\documentclass[]{emulateapj}

\usepackage{natbib}
\usepackage{bm}
\newcommand{\eqref}[1]{(\ref{#1})}

\shorttitle{}
\shortauthors{}

\begin{document}

\title{Interplay among Cooling, AGN Feedback and Anisotropic Conduction in the Cool Cores of Galaxy Clusters}

\author{H.-Y.\ Karen Yang\altaffilmark{1,2},
and Christopher S.\ Reynolds\altaffilmark{2}} 
\altaffiltext{1}{Einstein Fellow}
\altaffiltext{2}{University of Maryland, College Park, Department of Astronomy and Joint Space Science Institute} 
\email{Email: hsyang@astro.umd.edu}

\begin{abstract}

Feedback from the active galactic nuclei (AGN) is one of the most promising heating mechanisms to circumvent the cooling-flow problem in galaxy clusters. However, the role of thermal conduction remains unclear. Previous studies have shown that anisotropic thermal conduction in cluster cool cores (CC) could drive the heat-flux driven buoyancy instabilities (HBI) that re-orient the field lines in the azimuthal directions and isolate the cores from conductive heating from the outskirts. However, how the AGN interacts with the HBI is still unknown. To understand these interwined processes, we perform the first 3D magnetohydrodynamic (MHD) simulations of isolated CC clusters that include anisotropic conduction, radiative cooling, and AGN feedback. We find that: (1) For realistic magnetic field strengths in clusters, magnetic tension can suppress a significant portion of HBI-unstable modes and thus the HBI is either completely inhibited or significantly impaired, depending on the unknown magnetic field coherence length. (2) Turbulence driven by AGN jets can effectively randomize magnetic field lines and sustain conductivity at $\sim 1/3$ of the Spitzer value; however, the AGN-driven turbulence is not volume-filling. (3) Conductive heating within the cores could contribute to $\sim 10\%$ of the radiative losses in Perseus-like clusters and up to $\sim 50\%$ for clusters twice the mass of Perseus. (4) Thermal conduction has various impacts on the AGN activity and ICM properties for the hottest clusters, which may be searched by future observations to constrain the level of conductivity in clusters. The distribution of cold gas and the implications are also discussed.

\end{abstract}

\keywords{conduction --- galaxies: active --- galaxies: clusters: intracluster medium --- instabilities --- magnetohydrodynamics --- methods: numerical }


\section{Introduction}


Galaxy clusters with short central cooling times, or cool-core (CC) clusters, are predicted to host massive cooling flows \citep{Fabian94}, but the large amounts of cold gas and stars expected from this model are not observed \citep[e.g.,][]{Cardiel98, Peterson03, Sanders08}. This implies that the radiative losses in cluster cores are balanced by some heating mechanisms, the most promising among which is feedback from active galactic nuclei (AGN) \citep[see][for a review]{McNamara07}. This scenario is supported by the occurrence of AGN jet-inflated X-ray cavities within all CC clusters, and the fact that the cavity power (which approximates the AGN jet power) correlates with the core X-ray luminosity, suggesting a feedback cycle is at work \citep{Rafferty06}. Indeed, theoretical models incorporating kinetic AGN feedback into hydrodynamic, cooling intracluster medium (ICM) atmospheres have successfully achieved self-regulation and reproduce gross properties of CC clusters, in particular, the positive temperature gradients within the CCs \citep{Gaspari11, Li14, Prasad15}. 

However, the AGN/ICM system has rich physics not captured by these cooling, hydrodynamic, kinetic-feedback models.  Two additional pieces of physics are thought to be particularly important in the physics of CCs and will the subject of this paper.  Firstly, thermal conduction has the potential to be extremely important for the thermodynamic state of a CC \citep[e.g.][]{Voigt04,Voit11}.   While it cannot by itself offset the radiative cooling in a stable manner \citep{Stewart84, Bregman88, Soker03, Zakamska03, Voigt04, Pope06}, it may provide part of the heating, reducing the burden on the AGN, and can aid in the isotropization of the AGN heating throughout the core.  Secondly, the ICM is known to be magnetized with $\beta\sim 100$ \citep{Carilli02}, where $\beta$ is defined as the ratio between gas thermal pressure and magnetic pressure.  Under conditions characterizing the ICM, conductive heat fluxes flow locally along magnetic field lines and are strongly suppressed across field lines \citep{Spitzer62}.  Thus, the ICM must really be considered a magneto-hydrodynamic (MHD) system with conductive heat fluxes dictated by field geometry as well as temperature gradients \citep{Chandran98, Narayan01}.  

From a theoretical perspective, the real complexity of a CC arises from interactions between the kinetic feedback, radiative cooling, magnetic field evolution, and (anisotropic, magnetically-channeled) thermal conduction.  Magnetic fields may increase the structural integrity of jet-blown cavities, especially if enhanced by the action of magnetic draping as the cavity buoyantly rises \citep{Dursi08}.  This may, in turn, promote the driving of turbulence in and subsequent heating of the ICM \citep[this possibility was speculated on in ][]{Reynolds15}.  Somewhat less intuitively, anisotropic thermal conduction renders the CC unstable to the heat-flux driven buoyancy instability \citep[HBI; ][]{Quataert08, Parrish09, McCourt11} which, in quiescent atmospheres with extremely weak fields, saturates by re-orientating the field lines perpendicular to the temperature gradients, wrapping them onto the spherical isotherms \citep{Parrish09, Bogdanovic09}.  Thus, it might seem that even powerful thermal conduction is self-limiting, although semi-local simulations by \cite{Avara13} and \cite{Kunz12} have shown that the saturation property of the HBI is altered in more realistic magnetic field strengths, leading to radial magnetic filaments that can be long-lived conduits for heat flux. \footnote{The works of \cite{Avara13} and \cite{Kunz12} attribute the formation of these filaments to conservation of vertical magnetic flux in their semi-local domains --- in essence, the conserved flux is concentrated into filaments that are HBI stable, with the remaining volume possessing a field that is driven to be orthogonal to the temperature gradient.  Whether the same filament formation process will occur in a global geometry, given that magnetic flux can in principle be expelled from a core, has not been explored and is one motivation for the work presented in this paper.}  It is possible, however, that the HBI is too fragile to be relevant.   \cite{Ruszkowski10} and \cite{Ruszkowski11a} have shown that (driven) volume filling ICM turbulence could overwhelm any HBI-driven dynamics, although whether this is true for turbulence that is self-consistently driven by kinetic AGN feedback has never been explored.  One can readily imagine that the buoyantly rising jet-blown cavities may stretch field lines in the radial direction, counteracting the field-line wrapping of the HBI and allowing conductive heat fluxes to flow.  In the extreme, as speculated by \cite{Bogdanovic09}, one could imagine CCs in which the role of kinetic feedback was purely to stir the ICM and maintain open magnetic channels for the conductive heat flux.  At the same time, thermal conduction may strongly suppress local thermal instabilities that are important for the condensation of cold gas that, ultimately, becomes the fuel for the AGN exerting the kinetic feedback. 

In this Paper, we present the first simulations of CCs that include the four crucial physical components : MHD, radiative-cooling, jet-mediated feedback, and anisotropic thermal conduction.  Previous works have examined models that include only a subset of these processes and hence miss the important ``cross terms'', some of which we have discussed above.  We also examine, again for the first time in an ab-initio simulation, the mass dependence of the thermal conduction assuming that it has the Spitzer-like scaling.  The main questions that we address are : (1) Does the HBI play any significant role once we consider realistic magnetic field strengths and the jet-driven dynamics?   (2) What is the relative importance of conductive heating and direct AGN heating, and how does this change as we consider clusters of different mass?  (3) What are the impacts of conduction on the condensation of cold gas and the subsequent fueling of the AGN?  What are the predicted observables of these processes for clusters of difference masses?

The structure of the paper is as follows. We describe the numerical techniques and simulation setups in \S~\ref{sec:method}. In \S~\ref{sec:results}, we first present results for the simulations of the HBI without AGN feedback (\S~\ref{sec:noAGN}) and with AGN feedback (\S~\ref{sec:HBIAGN}). We then show the amount of conductive heating versus AGN heating in \S~\ref{sec:heating} and explore the impacts of conduction in \S~\ref{sec:impact}. Parameter dependencies of the results are presented in \S~\ref{sec:param}. We discuss the distribution of cold gas found in our simulations in \S~\ref{sec:coldgas} and comparisons with previous works in \S~\ref{sec:comparison}. Our main conclusions are summarized in \S~\ref{sec:conclusion}. 


\section{Methodology}
\label{sec:method}

\subsection{Numerical techniques}

We perform 3D MHD simulations with anisotropic thermal conduction to follow the evolution of cluster CCs under the influence of radiative cooling and AGN feedback. The simulations are carried out using the adaptive-mesh-refinement (AMR) MHD code FLASH \citep{Flash, Dubey08}. The MHD equations are computed using the directionally unsplit staggered mesh (USM) solver \citep{Lee09, Lee13}. The USM algorithm in FLASH is based on a finite-volume, high-order Godunov scheme and is combined with a constrained transport method to ensure the magnetic fields are divergence free. The simulation box is 1\ Mpc on a side, centering on an idealized CC cluster with initial conditions given in \S~\ref{sec:IC}. In simulations without AGN feedback, the simulation domain is refined progressively on the central region with a radius of 100\ kpc and kept fixed throughout the simulations. With a maximum refinement level of 7 in AMR, the finest resolution element is 1.95\ kpc. For runs including AGN, the central 10\ kpc is initially refined to highest resolution and AMR is turned on for regions with steep temperature gradients. This ensures that the regions affected by the jets, turbulence, and shocks are well resolved. The diode boundary condition is used, which is similar to the outflow boundary condition but does not allow matter to flow into the domain. Radiative cooling is computed using \cite{SutherlandDopita} assuming 1/3 solar metallicity. 

The MHD equations that governs the system we are solving are
\begin{eqnarray}
&& \frac{\partial \rho}{\partial t} + \nabla \cdot (\rho {\bm v}) = 0,\\
&& \frac{\partial \rho {\bm v}}{\partial t} + \nabla \cdot \left( \rho {\bm v}{\bm v}- \frac{{\bm B}{\bm B}}{4\pi} \right) + \nabla p_{\rm tot} = \rho {\bm g},\\
&& \frac{\partial {\bm B}}{\partial t} - \nabla \times ({\bm v} \times {\bm B}) = 0, \label{eq:ind}\\
&& \frac{\partial e}{\partial t} + \nabla \cdot \left[ (e+p){\bm v} - \frac{{\bm B}({\bm B} \cdot {\bm v})}{4\pi} \right] \nonumber \\ 
&& = \rho {\bm v} \cdot {\bm g} - \nabla \cdot {\bm Q} - n_e^2 \Lambda(T),
\end{eqnarray}
where $\rho$ is the gas density, ${\bm v}$ is the velocity, ${\bm B}$ is the magnetic field, ${\bm g}$ is the gravitational field, ${\bm Q}$ is the conductive heat flux, $\Lambda(T)$ is the cooling function, $n_e=\rho/\mu_{\rm e} m_{\rm p}$ is the electron number density ($\mu_{\rm e} = 1.18$ is the mean molecular weight per electron for the ICM with a third solar metallicity), $e=0.5\rho v^2 + e_{\rm th} + B^2/8\pi$ is the total energy density, and $p_{\rm tot} = (\gamma -1)e_{\rm th} + B^2/8\pi$ is the total pressure from the gas and magnetic fields. An equation of state for ideal gas with $\gamma=5/3$ is assumed. 

Under the conditions that characterize the ICM, the conductive heat is effective along magnetic field lines and strongly suppressed perpendicular to the field lines. The anisotropic conductive heat flux can be written as 
\begin{equation}
{\bm Q} = - \chi {\bm b}({\bm b} \cdot \nabla) T,
\end{equation}
where $T$ is the gas temperature, and 
\begin{equation}
\chi \simeq 4.6\times 10^{-7} T^{5/2} {\rm erg\ cm^{-1}\ s^{-1}\ K^{-1}}
\label{eq:chi}
\end{equation}
is the thermal conductivity \citep{Spitzer62}. The ion component of the heat flux is ignored since it is smaller than the electron contribution by a factor of $(m_i/m_e)^{1/2} \approx 42$. In reality, the thermal conductivity is likely to be suppressed compared to the Spitzer value because of magnetic field structures that are unresolved in the simulations. Here we choose the unsuppressed value in order to explore the most optimistic case where thermal conduction is in effect at its maximal strength. 

We implemented anisotropic conduction following the approach of \cite{Sharma07}, which applies a monotonized central (MC) limiter to the conductive fluxes in order to prevent negative temperatures in the presence of steep temperature gradients. Verification tests of the implementation of this algorithm in FLASH can be found in \cite{Ruszkowski11b} and \cite{Yang12}. An upper limit is imposed on the conductive diffusivity, $\kappa \equiv \chi T/p < \kappa_{\rm max} = 5 \times 10^{31}\ {\rm cm^{2}s^{-1}}$, in order to prevent extremely small diffusive timesteps. The application of the ceiling can be justified by the fact that it only affects the region containing the AGN jets and jet-shocked ICM. While currently treated as hot gas in the simulations, the jets are likely composed of relativistic particles, for which the Spitzer conductivity does not apply. As for the shocked ICM, although the conductivity is high by face values, the actual heat flux from the shocked to ambient gas is suppressed because magnetic field is aligned with the shock surface due to magnetic draping \citep{Lyutikov06, Dursi08}. Indeed, we find that varying the ceiling does not significantly affect our main conclusions.  

In this work, the effects of pressure anisotropy (i.e., Braginskii anisotropic viscosity) are not included. In a weakly collisional plasma, \cite{Kunz11} have shown that anisotropic viscosity can impair the growth of HBI. Pressure anisotropy can also drive plasma micro-instabilities that trap ions in magnetic field fluctuations tangled on scales of their gyro-radii and therefore stifle the heat flux \citep{Schekochihin10, Kunz14}. However, it is currently unknown whether electrons, which are the main contributors to the conductivity, would be affected by similar plasma instabilities. Moreover, in cluster cores, where the gas is collisional and has a lower plasma $\beta$, the impacts of anisotropic viscosity and micro-instabilities are expected to be significantly attenuated  \citep{Kunz11, Kunz12}. For the above reasons, we choose to focus on thermal conduction alone and leave the investigation of anisotropic viscosity in future work. 


\subsection{Cluster initial conditions}
\label{sec:IC}

An idealized, magnetized CC cluster is set up in order to probe the interplay among radiative cooling, AGN feedback and anisotropic conduction. In the fiducial run, the ICM is initialized using an empirical temperature fit \citep{Pinzke10} to the surface brightness profile of the Perseus cluster \citep{Churazov03}:
\begin{equation}
T = T_{\rm 0}\ \frac{1+(r/r_{\rm 0})^3}{2.3+(r/r_{\rm 0})^3} \left[ 1+(r/r_{\rm 1})^2 \right]^{-0.32},
\end{equation}
where $T_{\rm 0} = 7$\ keV, $r_{\rm 0} = 71$\ kpc, and $r_{\rm 1} = 380$\ kpc. The density profile is computed assuming hydrostatic equilibrium with an NFW gravitational potential \citep{Navarro96}:
\begin{equation}
\Phi(r) = -\frac{GM}{r} \frac{\ln (1+r/r_{\rm s})}{\ln(1+c)-c/(1+c)},
\end{equation}
where $M$ is the cluster virial mass, $r_s \equiv r_{\rm vir}/c$ is the scale radius, $r_{\rm vir}$ is the virial radius, and $c$ is the concentration parameter. We adopt $M=8.5\times 10^{14}\ M_\odot$, $r_{\rm vir}=2.440$\ Mpc, and $c=6.81$ for the fiducial model. The gravitational potential is kept fixed throughout the simulations and self-gravity is ignored because the gas contributes little to the total gravity. 

Because of the strong temperature dependence of the thermal conductivity, we explore its effects for different cluster masses. Specifically, we run simulations for clusters with half and double of the mass of the Perseus cluster, $M=4.25\times 10^{14}\ M_\odot$ and $M=1.7 \times 10^{15}\ M_\odot$, respectively. For these simulations, the characteristic quantities (i.e., $T_{\rm 0}$, $r_{\rm vir}$, $r_{\rm 0}$, $r_{\rm 1}$, $c$) are scaled based on predictions from the self-similar solution \citep{Kaiser86} and numerical simulations (for $c$ of relaxed clusters, e.g., \cite{Meneghetti14}): $r \propto M^{1/3}$, $T \propto M^{2/3}$, and $c \propto M^{-0.08}$. As will be discussed in \S~\ref{sec:heating}, there are negligible differences in the conductive and non-conductive simulations for the cluster with $M=4.25\times 10^{14}\ M_\odot$. Therefore, for this mass we show only the AGN-only case when studying the mass dependences of the results. 

Radio observations indicate that clusters host large-scale, volume-filling diffuse magnetic fields of strength $0.1-10\ \mu$G with tangled geometry \citep[see][for a review]{Carilli02}. Although the magnetic field is not dynamically important ($\beta \sim 100$ for typical clusters), its orientation dictates the directions of various transport processes including thermal conduction. The structure of cluster magnetic field is complex and likely tangled; however, the characteristic scale of its tanglement (i.e., the coherence length $l_{\rm B}$) is still uncertain \citep[e.g.,][]{Vogt03, Kuchar11}. 

In our simulation setup, a tangled initial magnetic field is assumed for the ICM. The magnetic field strength is normalized such that $\beta \sim 100$ (i.e., the magnetic pressure drops with radius from cluster center in proportion to the thermal pressure). Since the coherence length of cluster magnetic fields is uncertain, we adopt $l_{\rm B}=100$\ kpc (for the fiducial run) and 10\ kpc. These values are chosen to represent cases where the coherence length is larger or smaller, respectively, than the critical length for HBI growth (see \S~\ref{sec:noAGN}). The method of initializing a tangled field is illustrated in \cite{Yang12} and \cite{Yang13}, to which we refer the readers for details. We note that the initial tangled field is not force free, and therefore a small, random velocity field is generated ($|{\bf v}|< 100\ {\rm km}\ {\rm s}^{-1}$) soon after the simulations start despite the velocities are set to zero initially. As discussed in \S~\ref{sec:comparison}. whether the cluster has existing perturbations may be important for the development of thermal instabilities (TI) and thus the early evolution of the cluster. However, once the cluster enters into the nonlinear phase after stirred up by multiple episodes of AGN feedback, the long-term evolution is no longer sensitive to the amount of initial perturbations. 


\subsection{SMBH accretion and feedback prescriptions}
\label{sec:AGN}

In this section we summarize important aspects of our prescriptions for the accretion and feedback of the central supermassive black hole (SMBH). Since resolving the scales of SMBH accretion disks is far beyond what current computational power can achieve, subgrid prescriptions with simplified assumptions are necessary to model the large-scale effects of the SMBH. The implementation of AGN feedback in FLASH is described in detail in \cite{Yang12b}. Here we give a brief summary and highlight the differences. 

SMBH accretion has typically been modeled in cosmological simulations assuming that the black hole accretes from the surrounding hot gas (i.e., the {\it hot} mode) with the Bondi-Hoyle-Lyttleton accretion rate:
\begin{equation}
\dot{M}_{\rm BH} \propto \dot{M}_{\rm Bondi} = 4\pi G^2 M_{\rm BH}^2 \rho / c_s^3, 
\label{eq:bondi}
\end{equation}
where $M_{\rm BH}$ is the mass of the SMBH, and $\rho$ and $c_s$ are the gas density and sound speed, respectively, at the Bondi radius $r_{\rm Bondi} \equiv GM/c_s^2$. The proportionality reflects the uncertainties in computing the Bondi accretion rate due to the fact that the Bondi radius is usually not resolved in these large-scale simulations. Earlier observation of \cite{Allen06}, which showed a correlation between the Bondi accretion rate and cavity power for some of the low-power jet systems, has been one of the primary motivations for the hot mode accretion. However, more recent analysis with more accurate jet power estimations by \cite{Russell14} have found a weaker correlation. Moreover, it is difficult for the Bondi accretion to explain higher-power jets, for which other mechanisms are required \citep[e.g.,][]{McNamara11}. As has been discussed by various authors, the simple Bondi accretion has many theoretical difficulties \citep[e.g.,][]{Pizzolato05, McNamara07, Gaspari11, Hobbs12, Yang12b, Gaspari13}. Despite these shortcomings, Bondi accretion cannot be ruled out as it could still play a role in some systems, particularly in low-power jets \citep{Russell14} or in systems where cold gas is depleted \citep{Gaspari15}.

An alternative mechanism is cold gas accretion \citep[i.e. the {\it cold} mode; ][]{Pizzolato05}. In a globally thermally stable system, cold, dense clumps of gas can condense out of the hot ICM via local TI when $t_{\rm TI} \lesssim 10\ t_{\rm ff}$, where $t_{\rm TI}$ is the timescale for TI to grow, and $t_{\rm ff}$ is the free-fall time \citep{McCourt12, Sharma12, Gaspari12, Li14, Meece15}. When this criterion is met as the cluster core cools down, cold clouds are precipitated and feed the central SMBH. This triggers AGN jets that heat the ICM above the threshold for clump formation before the cluster cools down again, thus closing the loop of AGN feedback. This scenario is supported by observations of atomic \citep{Crawford99, McDonald11, Werner14} and molecular \citep{Donahue00, Edge01, Salome06, Hamer14, Russell14} gas at the center of many CC clusters. Entropy profiles of observed CC clusters also appear to follow a baseline profile predicted by the precipitation-regulated feedback model \citep{Voit15}. However, we note that there exist discrepancies between the entropy profiles of the above-mentioned work \citep[based upon the work of ][]{Cavagnolo} and those derived from a different group \citep{Panagoulia14}. Moreover, it is still unresolved how exactly the cold gas is channeled to the SMBH and what the roles of angular momentum transport, self-gravity, and cloud-cloud collisions are \citep{Pizzolato10, Gaspari13}. Therefore, it is still premature to conclude whether cold mode feedback is the dominating SMBH accretion mechanism in clusters. 

Given the uncertainties, we adopt the code mode accretion in the fiducial run, but also explore the hot mode.  The hot accretion model, though not favored by current observational data, is included in this paper because it is still physically plausible. Moreover, the variation of the assumption of SMBH accretion is the dominant source of uncertainties in AGN subgrid models \citep{Yang12b} and thus is useful for testing the robustness of our results. For the hot mode, we compute the Bondi accretion rate (Eq.\ \ref{eq:bondi}) using quantities averaged within an accretion radius $r_{\rm accre}=4$\ kpc surrounding the central SMBH. We assume the proportionality in Eq.\ \ref{eq:bondi} to be a constant of unity since it is degenerate with the jet efficiency (see below). The mass of the SMBH is $3.4\times 10^8\ M_\odot$ \citep{Wilman05}. 

For the code mode, the accretion rate is estimated by \citep{Gaspari13, Li14} $\dot{M}_{\rm BH}=M_{\rm cold}/t_{\rm ff}$, where $M_{\rm cold}$ is the total amount of cold gas within $r_{\rm accre}=4$\ kpc and $t_{\rm ff}=5$\ Myr is the approximate free-fall time at $r_{\rm accre}$ in the Perseus cluster. Cold gas in our simulations is defined to be gas with temperature below $5\times 10^5$\ K and is dropped out from the hot gas \citep{Gaspari11} and replaced with a cold gas particle. The cold gas particles are passive tracers that follow the dynamics of the hot fluid and their masses are included in the accretion rate calculation when they enter the sphere of radius $r_{\rm accre}$. The amount of accreted mass within a simulation timestep, $\dot{M}_{\rm BH} dt$, is then subtracted from the mass of the cold gas particles within $r_{\rm accre}$ in a mass-weighted sense. This method of mass dropout allows us to save computational costs and focus on the evolution of the hot ICM, which is the main subject of this paper. The consequences of using this approach are discussed in detail in \S~\ref{sec:discussion}. We will show in \S~\ref{sec:param} that our results are not sensitive to the assumption about SMBH accretion and thus the treatment of cold gas in our simulations.

AGN feedback is modeled as bipolar jets following the approach of \cite{Yang12b}. Feedback is continuous for the hot mode accretion whereas for the code mode, it is triggered only when $\dot{M}_{\rm BH}$ is nonzero. When the AGN is active, mass, momentum, and energy are injected onto the grid as source terms in the hydrodynamic equations according to the rates
\begin{eqnarray}
\dot{M} &=& \eta \dot{M}_{\mathrm{BH}}, \nonumber \\
|\dot{P}| &=& \sqrt{2\eta \epsilon} \dot{M}_{\mathrm{BH}} c \label{eq:jets}\\
\dot{E} &=& \epsilon \dot{M}_{\mathrm{BH}} c^2, \nonumber
\label{eq:jet}
\end{eqnarray}
where $\eta$ is the mass loading factor and the injected energy is purely kinetic. Purely kinetic jet feedback, though having shortcomings \footnote{Purely kinetic jets tend to form radially elongated bubbles instead of `fat' bubbles as observed in Perseus. Note, however, that not all observed cavities are fat.}, is adopted because it has successfully reproduced many key observables of CC clusters \citep[e.g.,][]{Gaspari11}. Furthermore, the properties of the CCs are relatively insensitive to the fraction of energy injected as kinetic vs.\ internal energy \citep{Yang12b}. We choose a mass loading factor of $\eta=1$, which is motivated by mass conservation and the fact that little mass growth is expected for radio mode AGN \citep{Li14}. The fiducial value of the feedback efficiency $\epsilon$ is 0.1 and 0.001 for the hot and cold mode accretion, respectively, which are within the ranges suggested by previous hydrodynamic simulations of AGN feedback in cluster cores \citep{Gaspari11, Li14}. The feedback is applied to a cylinder with radius $r_{\rm ej}=2.5$\ kpc and height $2h_{\rm ej}=4$\ kpc. Jet precession with an angle of $\theta_p=15^\circ$ and a period of 10\ Myr is assumed for the fiducial case. We experimented cases with no jet precession and found similar results with previous studies \citep{Vernaleo06} that the jets tend to drill through the ICM and create a low-density channel along the jet axis. As a result, whether the AGN could prevent the cooling catastrophe while preserving the observed ICM characteristics is very sensitive to the feedback efficiency parameter $\epsilon$. Namely, when $\epsilon$ and thus the jet power is too large, long cavities extending far beyond the cooling radius would be seen even though the system would eventually self-regulate. To this end, for all the simulations shown in this paper, we followed previous works \citep{Gaspari11, Li14} and introduced a small-angle jet precession in order to improve the coupling between the AGN feedback energy and the ICM.  

The primary simulations discussed in this paper are summarized in Table \ref{tbl:param}. In section \S~\ref{sec:param} we will discuss the parameter dependencies of the results. Subgrid AGN models are simplistic representations of the complex SMBH accretion and feedback processes and necessarily involve various parameters, both physically (e.g., mode of accretion, feedback efficiency) and numerically (e.g., radius and height of jet nozzle, accretion radius). Because of the large computational expenses required by the conductive simulations, it is not feasible to perform simulations over the full parameter space. Fortunately, such parameter studies have been done by previous hydrodynamic simulations \citep[e.g.,][]{Sijacki07, Gaspari11, Yang12b, Li14a}. In particular, \cite{Yang12b} found that among all parameters in subgrid AGN models, results are most sensitive to the assumptions of the accretion model and the feedback efficiencies, which are what we choose to vary for our simulations. As will be shown in \S~\ref{sec:param}, the main conclusions of this paper do not depend sensitively on the details of the AGN subgrid model. For simulations of the HBI with no AGN, we stop the simulations at $t=1.5$\ Gyr, about 1.2 Gyr after the cooling catastrophe occurs. Simulations with AGN feedback, on the other hand, are run up to $t=3$ Gyr in order to establish a quasi-equilibrium state but at the same time to save the large computational costs required by the conductive simulations. We ran the fiducial cases (Run CA and Run A) to $t=6$ Gyr and verified that our main conclusions remain the same. 

\begin{table}[tp]
\caption{Simulations presented in this paper.}
\begin{center}
\begin{tabular*}{0.45\textwidth}{@{\extracolsep{\fill}} cccccc}
\hline
\hline
Name & Aniso.\ Cond.\ & AGN & $\beta$ & $l_{\rm B}$ (kpc) & $M_{\rm cl} (10^{15} M_\odot)$ \\ 
\hline
C1  & y & n & 100 & 100 & 0.85\\
C2 & y & n & $\infty$ & 100 & 0.85\\
C3 & y & n & 100 & 10 & 0.85\\
CA & y & y & 100 & 100 & 0.85\\
A & n & y & 100 & 100 & 0.85\\
CAM$h$ & y & y & 100 & 10 & 1.70\\
AM$h$ & n & y & 100 & 100 & 1.70\\
AM$l$ & n & y & 100 & 100 & 0.425\\
\hline
\hline
\end{tabular*}
\end{center}
\label{tbl:param}
\end{table}


\section{Results}
\label{sec:results}

\subsection{HBI with no AGN}
\label{sec:noAGN}

In this section, we present results from simulations including radiative cooling and anisotropic thermal conduction but no AGN feedback (Run C1-C3). We show that the strength and coherence length of the magnetic field can significantly impact the growth of HBI. Note that for these runs, without AGN feedback, anisotropic conduction alone is not enough to counteract radiative cooling and thus cooling catastrophe occurs at $t \gtrsim 0.3$\ Gyr. Also, previous simulations found that catastrophic cooling can be delayed when thermal conduction is included \citep[e.g.,][]{Bogdanovic09}. We do not find a significant delay because the adopted profiles for the Perseus cluster has a short cooling time and flat temperature profile near the center.

\begin{figure}[tbp]
\begin{center}
\includegraphics[scale=0.55]{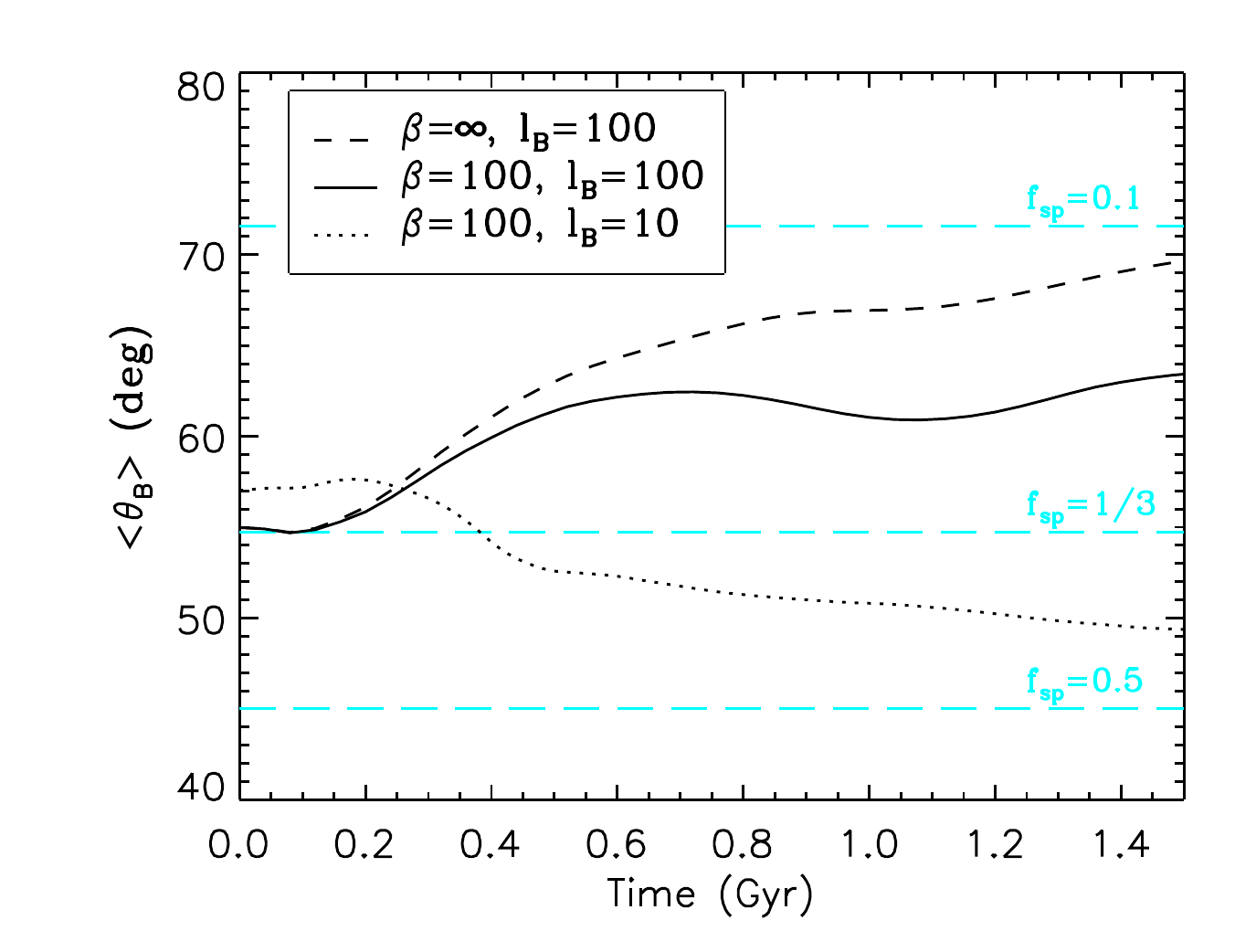} 
\caption{Evolution of the magnetic field orientation for the simulations including anisotropic conduction but no AGN feedback (Run C1-C3). Shown in the y-axis is the angle between the magnetic field and the radial direction ($\theta_{\rm B} \equiv \cos^{-1}|\hat{b} \cdot \hat{r}|$) averaged within $r=100$\ kpc. Different lines represent results for different plasma beta ($\beta$) and magnetic field coherence length ($l_{\rm B}$, in units of kpc). The horizontal lines show the effective Spitzer fraction $f_{\rm sp}=0.1$, 1/3, and 0.5 from top to bottom. This figure demonstrates that the HBI can be significantly impaired when a realistic $\beta$ in clusters is used and if the magnetic field is perturbed on scales smaller than $l_{\rm crit}$ (Eq.\ \ref{eq:lcrit}). }
\label{fig:thetaB}
\end{center}
\end{figure}

Figure \ref{fig:thetaB} shows the angle between the magnetic field and the radial direction ($\theta_{\rm B} \equiv \cos^{-1}|\hat{b}\cdot \hat{r}|$) averaged within $r=100$\ kpc. The solid and dashed lines are results from simulations with $l_{\rm B}=100$\ kpc but for two different values of $\beta$. Initially the magnetic field orientations are random, so $\langle \theta_{\rm B} \rangle \sim 55^\circ$, which corresponds to an effective Spitzer fraction $f_{\rm sp} \approx (\hat{b} \cdot \hat{r})^2 = \cos^2{\theta_{\rm B}} = 1/3$. As expected, for $\beta=\infty$, $\langle \theta_{\rm B} \rangle$ grows from an initial value of $55^\circ$ to $70^\circ$ at $t\sim 1.5$\ Gyr because the original tangled magnetic fields are re-oriented in the azimuthal directions by the HBI. On the contrary, when a more realistic plamsa beta $\beta=100$ is used, $\langle \theta_{\rm B} \rangle$ only grows from $55^\circ$ to $63^\circ$ at $t\sim 1.5$\ Gyr. The effective Spitzer fraction is about 0.12 and 0.21 at $t \sim 1.5$\ Gyr for $\beta=\infty$ and $\beta=100$, respectively. The rate of field-line re-orientation is much slower than the $\beta=\infty$ case, implying the HBI is suppressed when $\beta$ is small.  

The dependence of HBI saturation properties on the magnetic field strength was noted by \cite{Avara13}. They found that for moderate plasma beta values (which, for their local/small-box simulations, means $\beta \sim 10^7$), the saturation state of the HBI is qualitatively different from the standard HBI results ($\beta>10^9$), featuring magnetic field `filaments' that sustain heat fluxes at $\sim 10\%-25\%$ of the Spitzer value. These filaments are regions where magnetic tension is enhanced due to compressed field lines during the process of field-line wrapping and is enough for resist further growth of the HBI. Specifically, magnetic field perturbations are stable to HBI if the following condition is satisfied (Eq.\ 18 in \cite{Avara13} with recovered physical units, assuming $k_\perp \approx k_\parallel$ for an average perturbation):
\begin{equation}
-\frac{1}{2} \frac{g}{T^2} \frac{dT}{dr} + \frac{k_{\rm B}}{\mu m_{\rm p}} \frac{k^2}{\beta} > 0,
\end{equation}     
where $g$ is the amplitude of gravitational acceleration, $k_{\rm B}$ is the Boltzmann constant, $\mu$ is the mean molecular weight, $m_{\rm p}$ is the mass of a proton, and $k$ is the wave number of the perturbation. This equation illustrates the competition between the effects of a destabilizing temperature gradient and the stabilizing influence of magnetic tension. For a given system where $g$, $T$, $dT/dr$, and $\beta$ are known, there exists a critical length $l_{\rm crit}$ below which perturbations are HBI stable.  

For conditions in the Perseus cluster, we find that 
\begin{eqnarray}
l_{\rm crit} &\simeq& 84\ {\rm kpc} \left( \frac{\beta}{200} \right)^{-1/2} \left( \frac{g}{10^{-8}\ {\rm cm\ s^{-1}}} \right)^{1/2} \nonumber \\ 
&& \times \left( \frac{T}{4\ {\rm keV}} \right) \left( \frac{dT/dr}{4\ {\rm keV}/100\ {\rm kpc}} \right)^{-1/2}.
\label{eq:lcrit}
\end{eqnarray}
Therefore, magnetic field perturbations with wavelengths greater than $l_{\rm crit}$ would be subject to the HBI, while fluctuations on scales smaller than $l_{\rm crit}$ would be stable to the HBI because the magnetic tension itself is large enough to resist the growth of perturbations. The fact that $l_{\rm crit}$ is at a comparable scale to the radius of the CC means that a substantial portion of the HBI modes is suppressed for realistic ICM parameters. As a result, magnetic field lines are not completely wrapped as expected for the standard HBI simulations with high $\beta$. Instead, there are channels of field lines that are not perpendicular to the temperature gradient and are able to conduct heat. This is why the conductivity is maintained at a level of $\sim 22\%$ of the Spitzer value for the $\beta=100$ case. We note, though, that Eq.\ \ref{eq:lcrit} is derived from a local linear perturbation analysis, and thus the value of $l_{\rm crit}$ quoted here is only a crude estimate and can vary by a factor of order unity. Regardless, the main point is that HBI should be much suppressed under a scale of tens of kpc for realistic $\beta$ in clusters. 

\begin{figure}[tbp]
\begin{center}
\includegraphics[scale=0.3]{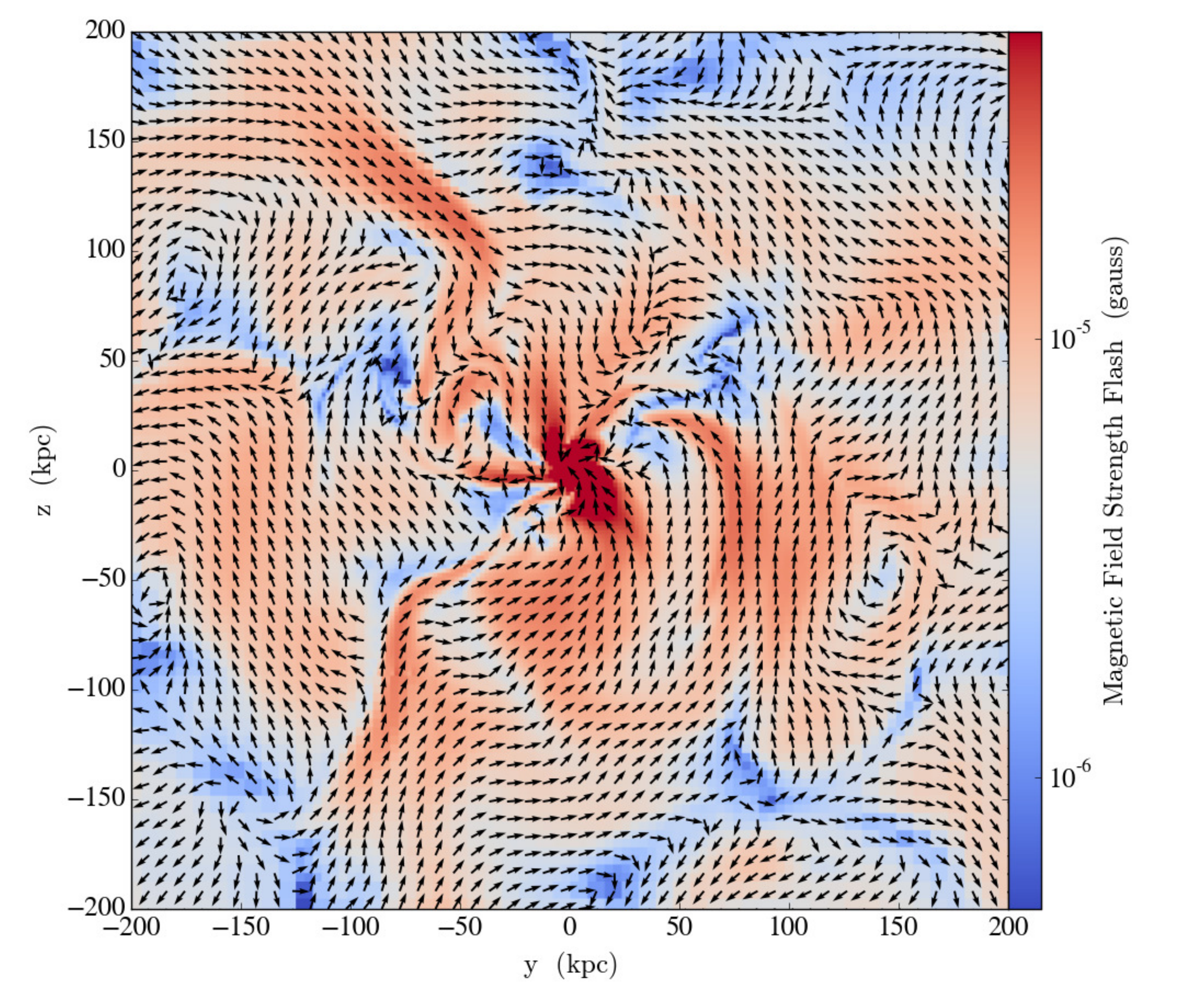} 
\caption{Magnetic field strengths and orientations (shown in vectors) for Run C1 ($\beta=100$, $l_{\rm B}=100$\ kpc) at $t=1$\ Gyr. Despite the overall field-line wrapping effect of the HBI, regions that are stable to the HBI appear as magnetic field `streams' with enhanced strengths and widths of tens of kpc, consistent with the expectation from Eq.\ \ref{eq:lcrit}. These streams can resume conduction to a level of $\sim 22\%$ of the Spitzer value (solid line in Figure \ref{fig:thetaB}).}
\label{fig:Bstream}
\end{center}
\end{figure}

Do we see magnetic field filaments as found in previous simulations \citep{Kunz12, Avara13}? Figure \ref{fig:Bstream} shows the field strength overplotted with vectors representing the directions of the magnetic field. We find magnetic `streams' as bands with larger field strengths and widths of tens of kpc, consistent with the above estimate. They do not look like `filaments' because their widths are much greater than the simulation resolution. In contrast, these HBI stable regions appear as narrow filaments in the previous simulations because the higher $\beta$ adopted leads to $l_{\rm crit}$ that is comparable to the size of their resolution elements. 

Equation \ref{eq:lcrit} implies that if the initial magnetic field is perturbed {\it only} on scales below $l_{\rm crit}$, the HBI should not develop at all. We tested this hypothesis using a simulation which starts from an initial field of $l_{\rm B}=10$\ kpc and $\beta=100$ (Run C3, dotted line in Figure \ref{fig:thetaB}). The figure shows that indeed, unlike the $l_{\rm B}=100$\ kpc case, the HBI has no effect and $\langle \theta_{\rm B} \rangle$ does not grow with time. The value of $\langle \theta_{\rm B} \rangle$ even decreases because the cooling catastrophe near the cluster center pulls the magnetic field into the radial direction after $t\sim 0.3$\ Gyr, similar to the inner $\sim 20$\ kpc shown in Figure \ref{fig:Bstream} for Run C1. We have also verified this result using clusters with longer cooling times in order to isolate the effect of $l_{\rm B}$ from the radial bias induced by the cooling catastrophe (see Appendix \ref{appendix}).

Rotation measure observations through AGN radio lobes have shown cluster magnetic fields tangled on kpc scales \citep{Carilli02}. However, whether there is a turnover in the magnetic field power spectrum is difficult to determine from current data and is uncertain \citep[e.g.,][]{Vogt03, Kuchar11}. If future observations find that a small magnetic field coherence length (compared to $l_{\rm crit}$) is typical, then the HBI should be unimportant and no field-line wrapping is expected. Even if observations suggest a large $l_{\rm B}$, the effects of the HBI should still be significantly attenuated because of the suppressed small-scale modes by magnetic tension. 


\subsection{HBI with AGN}
\label{sec:HBIAGN}

In the following, we compare the simulations with $\beta=100$, $l_{\rm B}=100$\ kpc with and without AGN feedback, i.e., Run CA and C1, respectively. We investigate whether AGN jet-driven turbulence is volume-filling and how much it can change the magnetic field topology and compete with the field-line wrapping effect of the HBI. 

\begin{figure}[tbp]
\begin{center}
\includegraphics[scale=0.55]{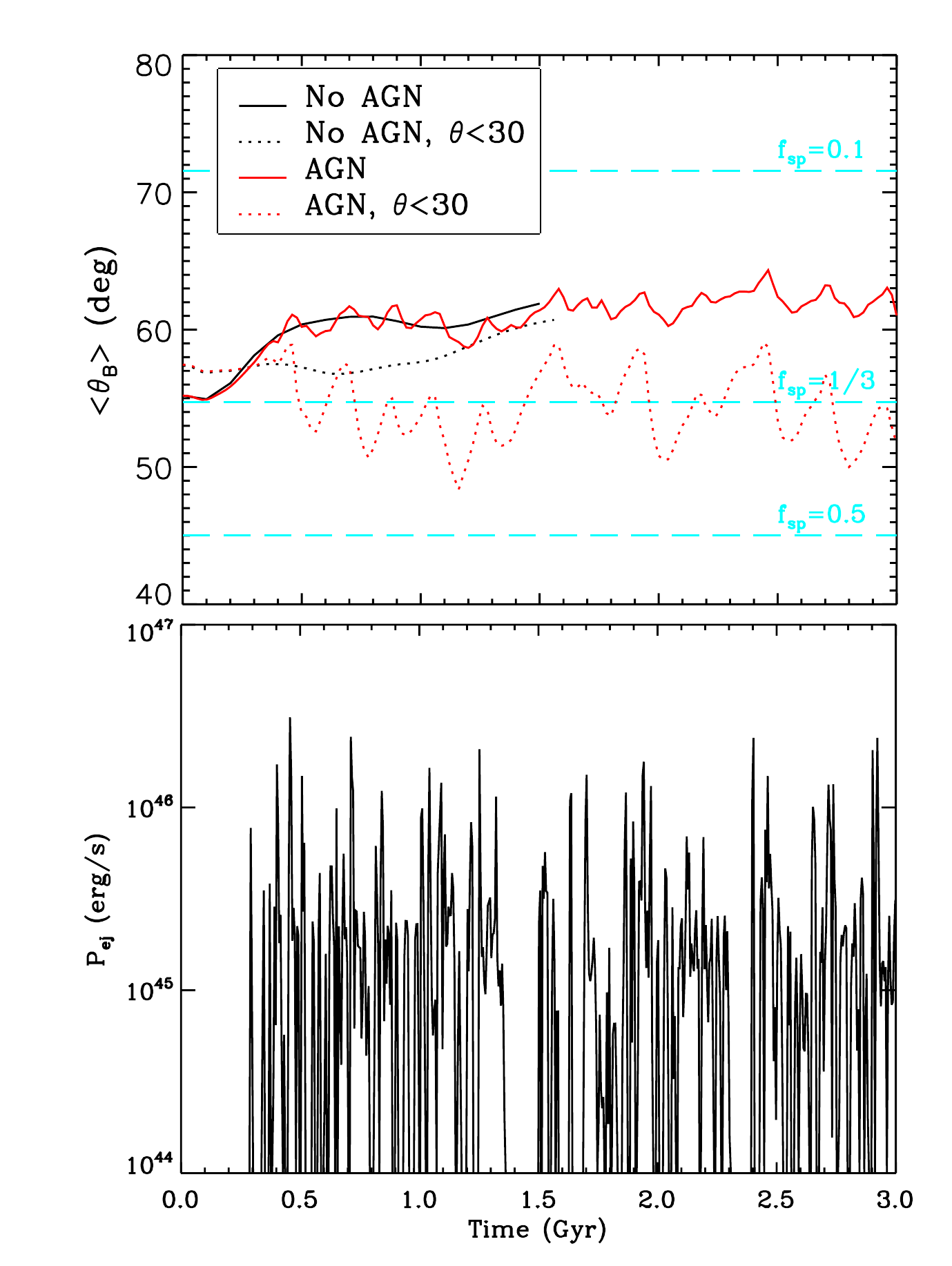} 
\caption{{\it Top}: The evolution of $\langle \theta_{\rm B} \rangle$ for runs with (red, Run CA) and without (black, Run C1) AGN feedback. The solid curves are for $\theta_{\rm B}$ averaged within $r =100$\ kpc, and the dotted lines show the same averaged $\theta_{\rm B}$ but only for regions $30^\circ$ away from the jet axes (i.e., within two cones). The definitions of $\theta_{\rm B}$ and the horizontal lines are the same as in Figure \ref{fig:thetaB}. {\it Bottom}: AGN jet power vs. time for Run CA.}
\label{fig:thetaB_AGN}
\end{center}
\end{figure}

Figure \ref{fig:thetaB_AGN} (top panel) shows the evolution of $\langle \theta_{\rm B} \rangle$ for runs with (red lines) and without AGN feedback (black lines). The solid curves are for $\theta_{\rm B}$ averaged within a sphere with radius of 100\ kpc, and the dotted lines show the averaged $\theta_{\rm B}$ only for regions $30^\circ$ away from the jet axes (i.e., within two cones). The jet power versus time for Run CA is plotted in the bottom panel.  In contrast to the case without AGN, the evolution of $\langle \theta_{\rm B} \rangle$ exhibits dips that are temporally correlated with large AGN outbursts, implying the HBI is counteracted by the field-line stirring effect of the AGN. However, the influence of the AGN is directional. For regions along the jet axis, the field lines are stirred up enough so that the effective Spitzer fraction is maintained at $\sim 1/3$ (red doted line). This effect is weaker when all angles are considered (red solid line) because this region only occupies a small volume within the core. 

\begin{figure*}[tbp]
\begin{center}
\includegraphics[scale=0.6]{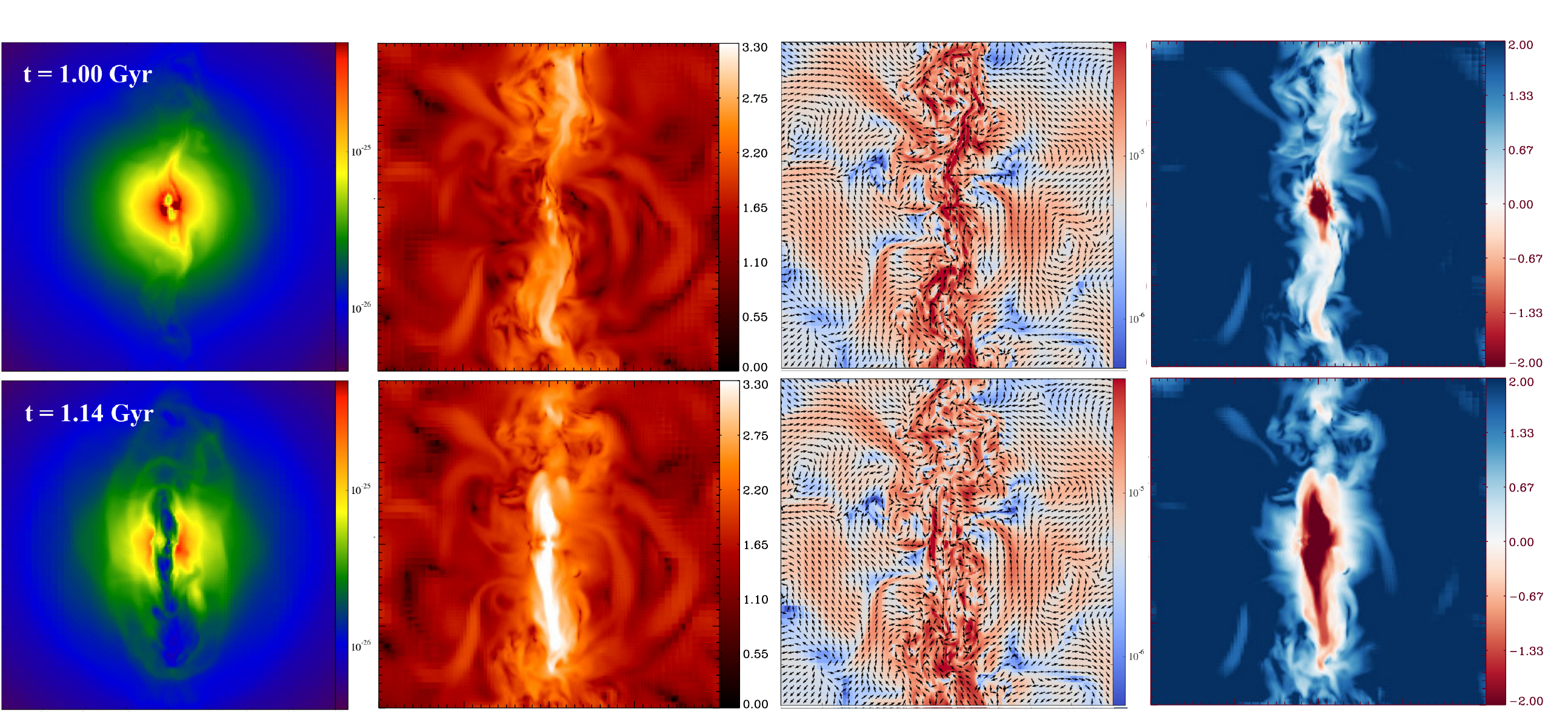} 
\caption{From left to right are slices of gas density (in units of ${\rm g\ cm^{-3}}$), the incompressible component of gas velocity (in units of ${\rm km\ s^{-1}}$ in logarithmic scale), magnetic field overplotted with vectors pointing at its directions (in units of Gauss), and Richardson number (in logarithmic scale; see the text for definition) at $t=1$\ Gyr (top row) and $t=1.14$\ Gyr (bottom row) for Run CA. Each panel shows a 200 kpc $\times$ 200 kpc region centering on the cluster center. Regardless the stage of AGN activity, the magnetic field orientations are randomized by turbulence generated along the jet axis, restoring conductivity to $\sim 1/3$ of the Spitzer value (red dotted line in the top panel of Figure \ref{fig:thetaB_AGN}).}
\label{fig:AGNslices}
\end{center}
\end{figure*}

The impact of AGN feedback on the magnetic field structure is further illustrated by the simulation snapshots shown in Figure \ref{fig:AGNslices}, which shows slices of gas density, the incompressible component of gas velocity \footnote{We decompose the velocity field into its compressible and incompressible components, ${\bf v} = {\bf v}_{\rm c} + {\bf v}_{\rm i} = \nabla \psi + \nabla \times {\bf a}$ and obtain the scalar field $\psi(r)$ by solving $\nabla^2 \psi = \nabla \cdot {\rm v}$ using the standard Fast Fourier Transform methods. This technique allows us to separate the compressible component, which mainly composes of weak shocks and sound waves, from the incompressible component, which includes turbulence, shear in bulk flows, and $g$-modes.} (which mainly measures the turbulent velocity and shear in bulk flows), the magnetic field strength, and the Richardson number (see definition below) at $t=1$\ Gyr (top row) and $t=1.14$\ Gyr (bottom row). At $t=1$\ Gyr, the whole cluster core is relatively relaxed and the AGN has just started the next cycle of large outbursts and inflated two cavities in the central $r\lesssim 15$\ kpc (top left panel). At $t=1.14$\ Gyr, these cavities have buoyantly risen beyond 100 kpc and another pair of cavities can be seen at a distance of $r\sim 40$\ kpc from the cluster center (bottom left panel). \footnote{It is worth clarifying that the bottom left panel in Figure \ref{fig:AGNslices} represents multiple pairs of buoyantly rising bubbles, instead of a low-density channel penetrated by a single, over-pressurized jet, which we verified using a map of gas pressure.} Despite the distinct states of AGN activity and ICM morphology, the underlying velocity (second column) and magnetic field (third column) structures are in fact quite similar, featuring a turbulent zone within $\sim 30^\circ$ from the jet axis and an ambient region where the ICM is largely undisturbed. Within the `cones' of influence of the AGN, turbulence is generated by vortices as the bubbles are disrupted and mixed with the ambient ICM. Because of repeated AGN events, the turbulent velocity is sustained at a few hundreds of ${\rm km\ s^{-1}}$ even when the AGN is quiescent. The magnetic field orientations are randomized by the turbulence, resulting in a Spitzer fraction of $\sim 1/3$ (red dotted line in the top panel of Figure \ref{fig:thetaB_AGN}). The main difference of the velocity and magnetic field structures between the two snapshots is the enhanced velocities and radially-stretched field lines in the wakes of the more recently formed bubbles. The radial bias of magnetic field trailing behind these previously inflated cavities causes the dips in $\langle \theta_{\rm B} \rangle$ that are slightly lagged the moments of AGN injections (see Figure \ref{fig:thetaB_AGN}). 

More quantitatively, the ability for gas shear motions to overcome the HBI could be estimated from the Richardson number \citep{Turner73, Ruszkowski10}, $R_i \equiv gr(d\ln T/d\ln r)/\sigma^2$, which is the ratio of the restoring buoyant force in a magnetized medium to the inertial term ($\rho {\bm v}\cdot \nabla {\bm v}$), and $\sigma$ is the turbulent velocity. For systems with $R_i \gg O(1)$, buoyancy is dominant; for $R_i \ll O(1)$, there is sufficient kinetic energy to make the velocity field isotropic. The values of the Richardson number at $t=1$ and $t=1.14$\ Gyr are plotted in the rightmost column in Figure \ref{fig:AGNslices}. We find that the region where magnetic field lines are tangled is coincident with gas with $R_i \lesssim O(1)$, demonstrating that the level of turbulence generated by the AGN is sufficient to overcome the HBI and randomize the velocity field. Note that the transition between buoyancy and flow gradient dominated regimes occurs at a critical Richardson number, $R_{i_c}$. For hydrodynamic flow, $R_{i_c} \sim 1/4$; however, order unity differences may exist for the MHD case.

For the ambient region (i.e., $\gtrsim 30^\circ$ from the jet axis), we find little influence of the AGN on the {\it incompressible} velocity field \footnote{AGN {\it do} have impacts on the {\it compressible} velocity field as they generate weak shocks and sound waves.} and the magnetic field: they are essentially consistent with the initially imposed fluctuations. This implies that little turbulence is generated by the HBI itself \citep{Bogdanovic09, McCourt11} or by the decay of $g$-modes. The latter may be due to the following reasons. First, $g$-modes are excited only when the condition $w<w_{\rm BV}$ is met, where $w$ is the characteristic frequency of the driving force and $w_{\rm BV}=(g/r)(d\ln T/d\ln r)$ is the Brunt-Vaisala frequency in a magnetized medium \citep[e.g.,][]{Ruszkowski10}. For the parameters adopted in our simulations, $w_{\rm BV}\sim 3\times 10^{-16}\ {\rm s}^{-1}$ at $r\sim 100$\ kpc and decreases inward. Therefore, only driving forces with periods larger than $\sim 100$\ Myr can meet the above criterion. The typical recurrence time of AGN outbursts in our simulations, which is self-consistently determined from the self-regulated feedback and is also consistent with observed values, is $\sim 0.1-10$\ Myr \citep[][and references therein]{McNamara07}, while events with intervals longer than 100\ Myr are relatively rare (bottom panel in Figure \ref{fig:thetaB_AGN}). Moreover, unless $w$ is much smaller than $w_{\rm BV}$, the excited $g$-waves would only be trapped in a small range in radii around $r=100$\ kpc. Consequently, it is difficult for the AGN to generate volume-filling turbulence within the entire cluster core.    

Our MHD simulations confirm the findings of previous hydrodynamic simulations that AGN-driven turbulence is local and non-volume filling \citep{Scannapieco08, Heinz10, Gaspari11, Vazza13}. It is also in agreement with a recent study by \cite{Reynolds15}, which showed that the generation of volume-filling turbulence by AGN requires that the $g$-waves are not trapped only within a small area. If future observations, e.g., Astro-H, give evidences for volume-filling turbulence within cluster cores, other mechanisms may need to be invoked, such as $g$-mode excitation by galaxy motions \citep{Ruszkowski11a} and decay from large-scale turbulence generated by cluster mergers \citep[e.g.,][]{Heinz10, Vazza13}.     


\subsection{Conductive vs. AGN heating}
\label{sec:heating}

In the last section we have shown that AGN jet-driven turbulence can randomize the magnetic field lines and restore the thermal conductivity to 1/3 of Spitzer. The next important questions to be addressed are whether this promotes significant heat transfer from outer radii to help balance radiative cooling and how much it is compared to the amount of AGN heating. Because thermal conductivity has a strong dependence on gas temperature (Eq.\ \ref{eq:chi}), we present results for the fiducial Perseus-like cluster (Run CA and Run A) and a cluster with double the mass (Run CAM$h$ and Run AM$h$). To probe the maximum strength of the effects of conduction, for the higher mass cases we choose $l_{\rm B}=10$\ kpc so that the conductive heat fluxes are not impeded by the HBI. 

\begin{figure}[tbp]
\begin{center}
\includegraphics[scale=0.55]{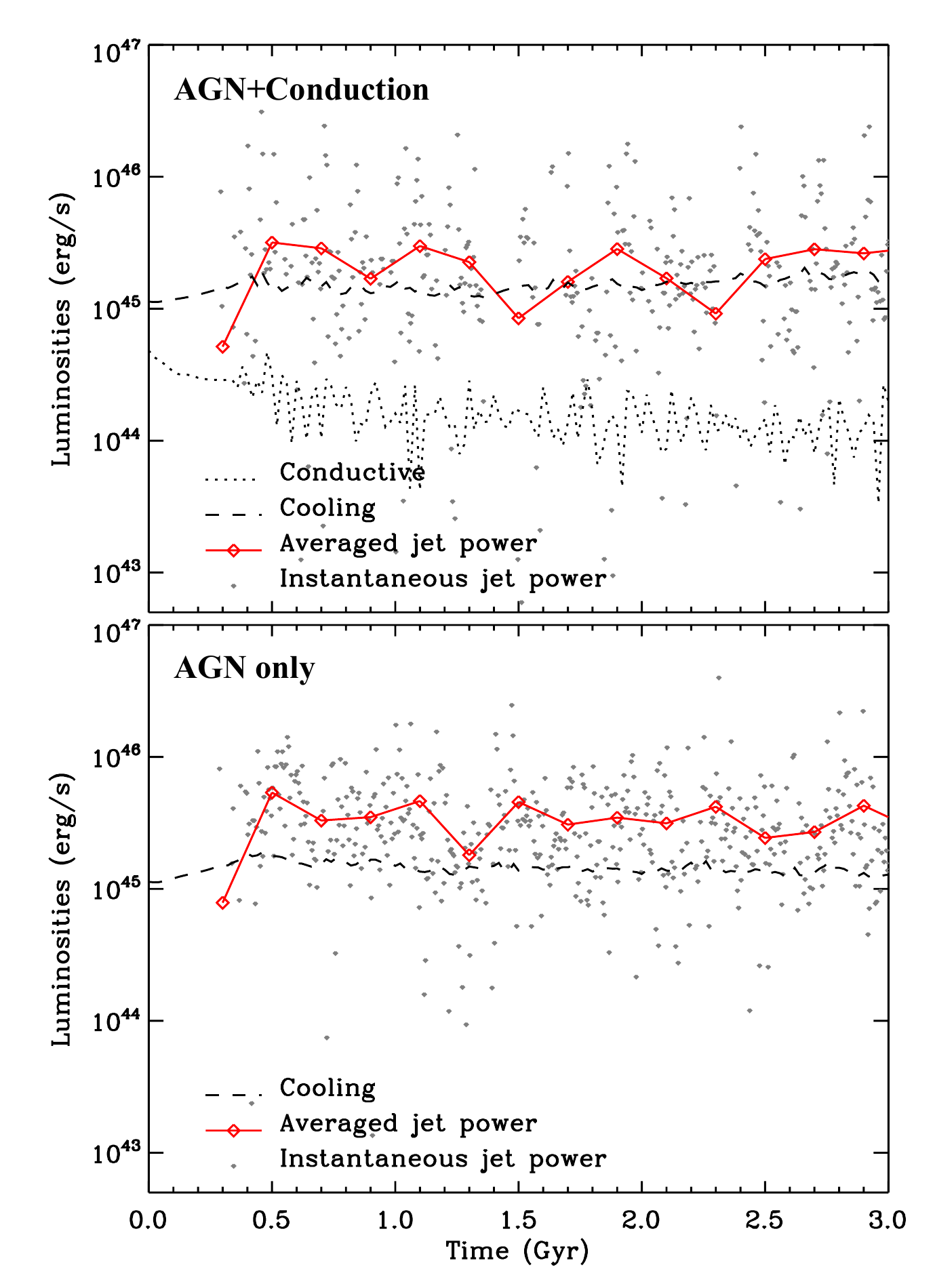} 
\caption{Heating and cooling luminosities for AGN simulations with (top, Run CA) and without (bottom, Run A) anisotropic conduction. The X-ray/cooling and conductive (shown in absolute values) luminosities are calculated within a cooling radius of $r_{\rm c}=100$\ kpc. The red curves show the AGN jet power averaged every 0.2 Gyr. For Run CA, after the cluster reaches a quasi-equilibrium state (i.e., $t\gtrsim1$\ Gyr), conductive heating contributes to $\sim 10\%$ of the radiative losses.}
\label{fig:heating}
\end{center}
\end{figure}

We quantify conductive heating within a sphere by the conductive luminosity, which is the conductive heat flux, 
\begin{equation} 
Q_{\rm cond}=-f_{\rm sp} \chi \partial T / \partial r,
\end{equation}
integrated across the surface of the sphere. We then compute the conductive luminosity within the cooling radius $r_{\rm c}=100$\ kpc, defined here as the radius at which the cooling time is equal to 3\ Gyr, and compare it to the X-ray luminosity within $r_{\rm c}$ and the AGN jet power. These quantities for Run CA are shown in the top panel of Figure \ref{fig:heating}. The results from a typical simulation of AGN feedback without conduction (Run A) are shown in the bottom panel for comparison.

For the standard AGN simulation (Run A), the cluster contracts due to radiative cooling until cold gas forms out of local TI when $t_{\rm c}/t_{\rm ff} \lesssim 10$ and triggers subsequent AGN activity after $t\sim 0.3$\ Gyr. Afterwards, the AGN self-regulates and the radiative cooling is balanced by AGN heating. The averaged jet power is greater than the X-ray luminosity within $r_{\rm c}$ because the bubbles can reach beyond $r_{\rm c}$ and thus the effective efficiency of transforming kinetic energy to thermal energy within $r_{\rm c}$ (referred to as `thermalization efficiency' hereafter) is $\sim 0.3$. 

For Run CA, the conductive luminosity starts from $\sim 50\%$ of the X-ray luminosity. As the cooling dominates, the cluster core again contracts and sets off the AGN. After the cluster reaches a quasi-static state after $t\sim 1$\ Gyr, the conductive heating only contributes to $\sim 10\%$ of the cooling losses, while the remaining is offset by AGN heating. We also computed the ratios between the conductive and X-ray luminosities for different radii and verified that conductive heating remains subdominant throughout the cluster core. The averaged jet power is only $\sim 2/3$ of that in the AGN-only simulation. The reasons are two-fold. First, the AGN does not need to inject so much energy with the aid of conduction. Second, because of the weaker jet power, the bubbles travel to shorter distances and thus the thermalization efficiency is higher than the AGN-only simulation.  

One might think the decrease in the amount of conductive heating is caused by the HBI. Indeed, the effective Spitzer fraction decreases from an initial value of 1/3 to $\sim 0.22$ after $t\sim 1$\ Gyr due to the HBI (red solid line in Figure \ref{fig:thetaB_AGN}). However, this is not enough for account for the degree of suppression of the conductive luminosity, implying some other factor is at play. To this end, we did an experiment and performed the same run as Run CA but with $l_{\rm B}=10$\ kpc, for which HBI should have no effects (for reason discussed in \S~\ref{sec:noAGN}). The conductive luminosity still drops to $\sim 2\times 10^{44}\ {\rm erg\ s^{-1}}$. What happens is that as the heat fluxes flow into the core region and cause the core temperature to rise. The temperature gradient in the core is reduced and thus subsequent conductive heat flux is inhibited. That is, conductive heating is a {\it self-limiting} mechanism. Though the relative importance of conductive to AGN heating obviously depends on cluster initial conditions and cluster masses (see below), we find that the it is a general trend for the conductive heat fluxes to decrease with time due to reduced temperature gradients. Moreover, this effect has more impacts on the long-term evolution of conductive heating than the temporarily-enhanced Spitzer fraction due to AGN-driven turbulence (note also that the conductive luminosity does not simply follow the evolution of $\langle \theta_{\rm B} \rangle$ because of the varying temperature gradients). 

\begin{figure}[tbp]
\begin{center}
\includegraphics[scale=0.55]{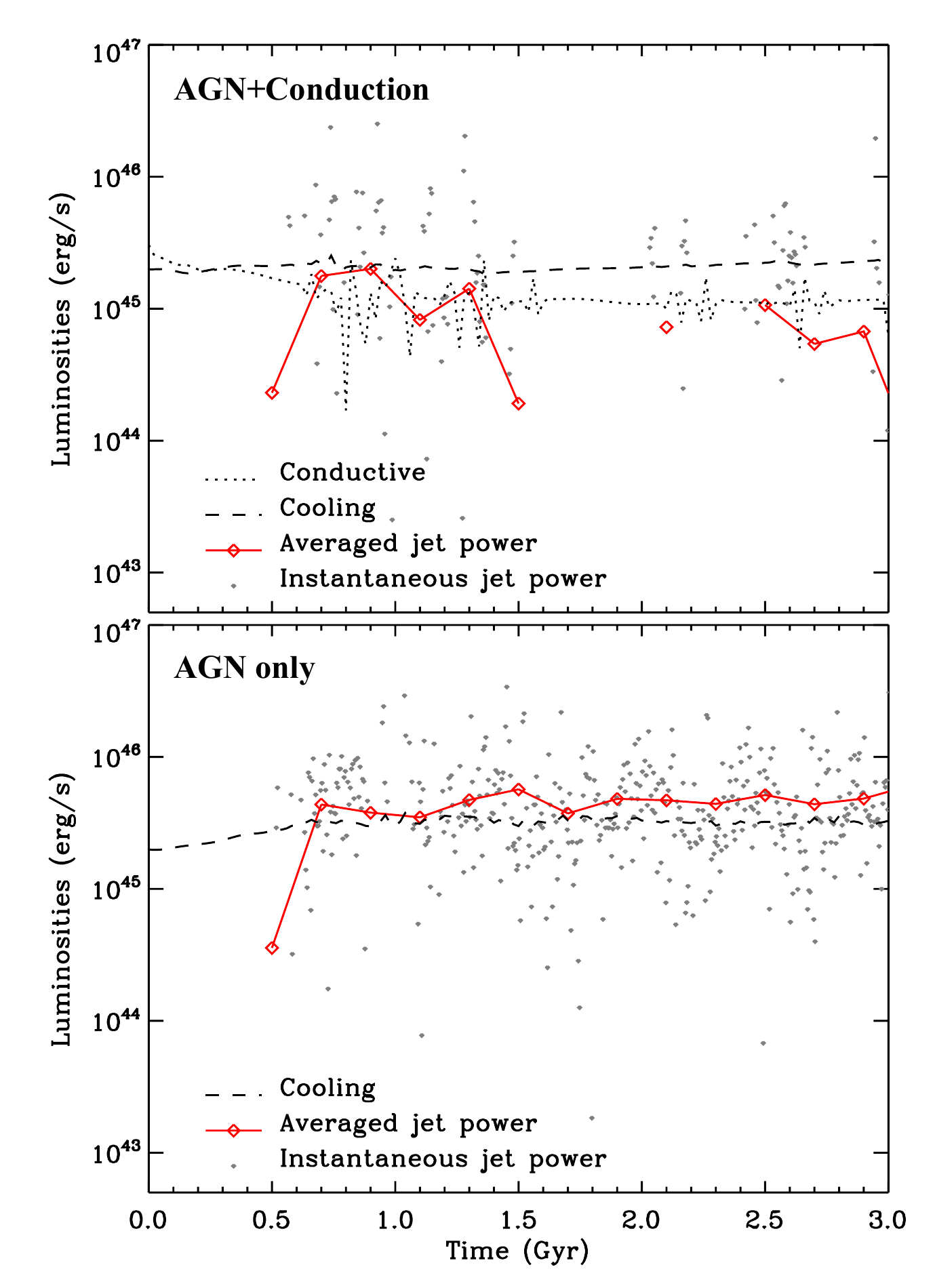} 
\caption{Same as Figure \ref{fig:heating} but for the high-mass cluster. Luminosity components for simulations with (Run CAM$h$) and without (Run AM$h$) conduction are shown in the top and bottom panels, respectively. For this cluster mass, conductive heating can offset $\sim 50\%$ of the radiative losses.}
\label{fig:heating_M15}
\end{center}
\end{figure}

The higher mass cases are presented in Figure \ref{fig:heating_M15}. Again, for the run without conduction (bottom panel), AGN feedback is self-regulated after $t\sim 0.7$\ Gyr, with the averaged jet power roughly balancing the X-ray luminosity within $r_{\rm c}=125$\ kpc, i.e., the thermalization efficiency within $r_{\rm c}$ is $\sim 100\%$. For the simulation including conduction (top panel), conductive heating is much more important than in the lower mass case, as expected. At the beginning of the simulation, it is even slightly larger than the cooling losses. However, the conductive luminosity decreases with time for the same self-limiting effect due to reduced temperature gradients (apparent from Figure \ref{fig:M15prf}) as discussed in the last paragraph. The AGN feedback is initiated at $t\sim 0.5$\ Gyr because at that point conduction alone is not able to withhold cooling. Afterwards, on average both the AGN and conductive heating contribute to $\sim 50\%$ of the radiative cooling losses. Contrasting with the AGN-only simulation, the AGN only needs to inject $\sim 1/4$ of feedback energy with the help of conduction, again due to weaker power of the jets and the resultant higher thermalization efficiency.   

We do not show the results for the lower-mass ($M=4.25\times 10^{13}\ M_\odot$) cluster here, because we find negligible differences between the simulations with and without conduction. Comparing the amount of conductive heating to the radiative losses, their ratio starts from $\lesssim 10\%$ at $t=0$ and decreases to a few percent level afterwards, causing little differences in all aspects.  

In summary, we find that conductive heating tends to be self-limited because of reduced temperature gradients itself introduces. This again reflects the dilemma of conduction being the main heating mechanism in CC clusters: conductive heat fluxes rely on temperature gradients, which they tend to erase \citep{Stewart84, Bregman88, Soker03, Pope06}. This effect is more significant than the enhanced Spitzer fraction enabled by AGN-driven turbulence in terms of long-term evolution of the cluster. For Perseus-like clusters, conductive heating is $\sim 10\%$ of the radiative cooing while AGN feedback is still the primary heating mechanism. For clusters doubling the mass of the Perseus, however, conductive heating can offset $\sim 50\%$ of the cooling losses. As a result, the AGN in these clusters do not need to operate in their full power. In principle, this could cause a shallower slope in the observed $P_{\rm cav}$--$L_{\rm X}$ relationship at the high-mass end ($P_{\rm cav}$ is the jet power estimated from cavity enthalpy and $L_{\rm X}$ is the X-ray luminosity). Indeed, the slope of the observed relation is less than unity \citep{Rafferty06, Birzan13}. However, it could also be that AGN are over-powered in lower mass systems. More importantly, systematic uncertainties in deriving the cavity power are substantial, mainly due to the unknown bubble ages. Therefore, a factor of 2 suppression of AGN power as suggested by our simulations would be very difficult to be detected observationally. We discuss other possible observable signatures of conduction in the next section.    


\subsection{Impact of conduction on AGN feedback}
\label{sec:impact}

We have shown that for the hottest clusters, thermal conduction can be as important as the AGN in terms of ICM heating. As evident from Figure \ref{fig:heating_M15}, conduction also impacts the pattern of AGN activity and therefore how the AGN interacts with the ICM. In this section we present detailed comparisons for simulations of AGN feedback with and without anisotropic conduction. We will focus on the hottest clusters (i.e., Run CAM$h$ and Run AM$h$) in order to explore the maximal effect of conduction, and we will also refer to the lower-mass cases (Run CA, Run A, and Run AM$l$) when discussing the mass dependence of the results. The goal is to understand the various influences of thermal conduction with the hope of providing clues of its observable signatures.

\begin{figure*}[tbp]
\begin{center}
\includegraphics[scale=0.7]{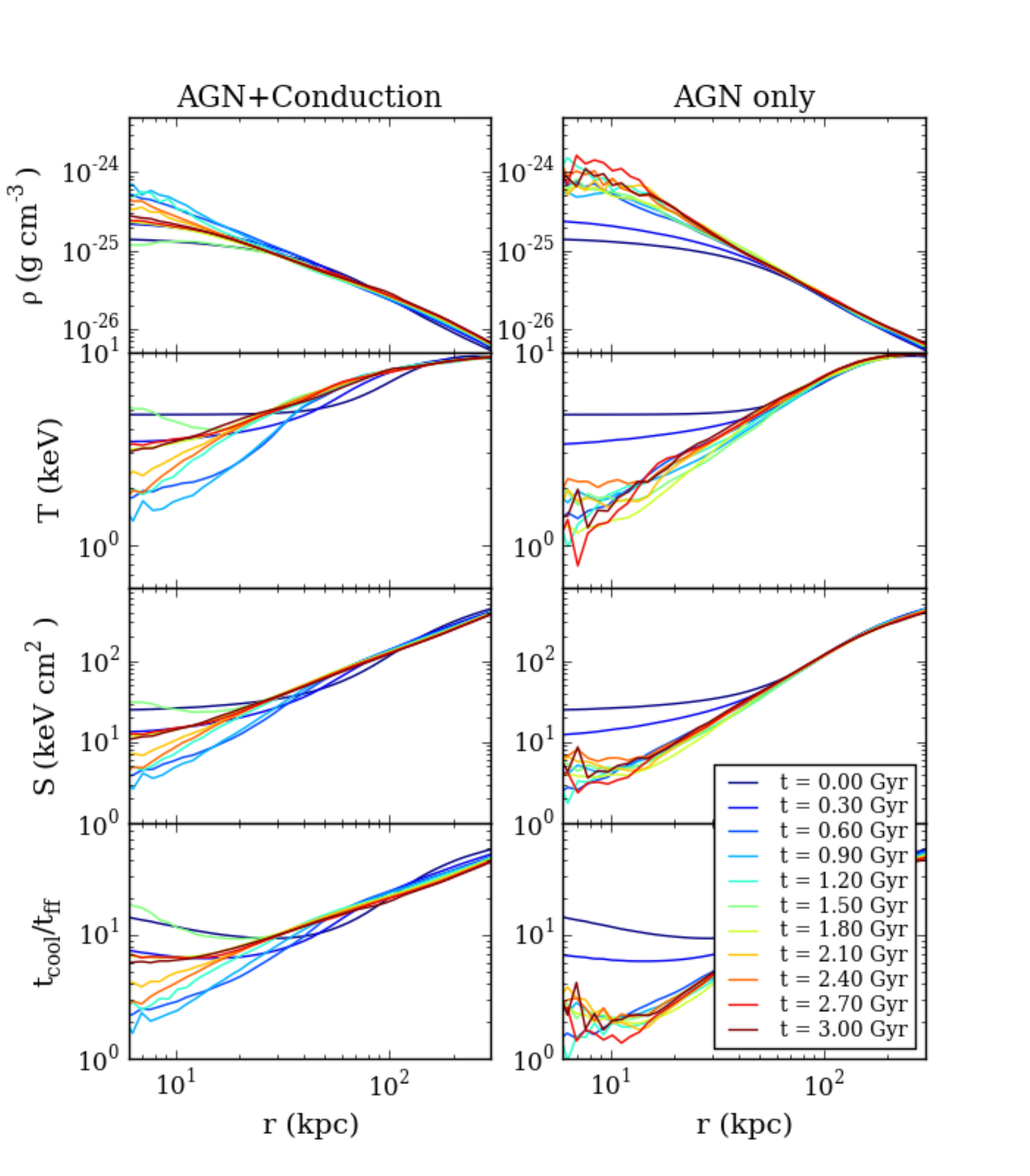} 
\caption{X-ray emission-weighted profiles of the high-mass cluster for the AGN simulations with (Run CAM$h$, left panels) and without (Run AM$h$, right panels) conduction. From top to bottom are profiles of gas density, temperature, entropy, and the ratio between cooling time and free-fall time. }
\label{fig:M15prf}
\end{center}
\end{figure*}

As discussed in the previous section, because of the contribution from conductive heating, the AGN jet power is on average smaller in the conductive simulations. These weaker jets deposit energy to a shorter distance from the cluster center, and therefore the heating is more efficient within this distance. As a result, the profiles of temperature, entropy, and $t_{\rm c}/t_{\rm ff}$ are more effectively elevated in the conductive case, while those for the AGN-only simulation in general stay close to their minimum (see Figure \ref{fig:M15prf}). The same trend is found for the Perseus-like cluster, though the differences between the conductive and non-conductive simulations are smaller.  

We note that for the high-mass cluster, even though conductive heating is important, the temperature profiles still preserve its CC appearance, without completely erasing the temperature gradients as one might naively expect. The entropy profiles were never elevated much above the critical balance between cooling and conduction \citep[see Figure 4 in][]{Voit11} so that the cluster never had an isothermal core. This is because the conduction has a slight dominance over cooling at larger radii, while allowing the inner core to be regulated by cooling and AGN feedback. At the same time, the deep gravitational potential of the massive cluster makes it difficult for AGN feedback to heat the core beyond the critical condition.   


Another prominent distinction of Run CAM$h$ and Run AM$h$ is the frequency of AGN outbursts (see the instantaneous jet power shown as grey dots in Figure \ref{fig:heating_M15}). With conduction, the AGN activity is episodic and has several quiescent phases (e.g., $t=1.6-2.0, 2.2-2.4, 2.7-2.9$\ Gyr), whereas without conduction the AGN activity is much more steady. The duty cycle, defined as the fraction of time when the AGN is active measured between $t=1$\ Gyr and $t=3$\ Gyr, is $4.4\%$ and $40.0\%$ for the runs with and without conduction, respectively. Consequently, multiple pairs of cavities are rare in the conductive simulation but quite common in the AGN-only case for these very hot clusters. For the Perseus-like cluster, on the other hand, there are not clear morphological distinctions (multiple pairs of cavities like the observed ones are common, see the bottom-left panel in Figure \ref{fig:AGNslices}), though the AGN duty cycle for the conductive case is again smaller ($38.6\%$) than for the non-conductive simulation ($52.2\%$). For comparison, the AGN in the lower-mass cluster (Run AM$l$) has a duty cycle of $58.4\%$. 

The infrequent AGN outbursts in the conductive simulations can be attributed to two effects. First, because of the elevated profiles, the condition required for local TI, $t_{\rm c}/t_{\rm ff}\lesssim 10$, is met less frequently. Moreover, conduction can suppress TI along magnetic fields \citep[e.g.,][]{Sharma10}. Although the Field length for the cold gas is unresolved in the simulations, that for the hot gas is, and therefore the growth of density perturbations in the hot ICM is still expected to be suppressed. Due to the above reasons, less cold gas is formed and thus less is accreted onto the central SMBH in the simulations including conduction.

\begin{figure}[tbp]
\begin{center}
\includegraphics[scale=0.55]{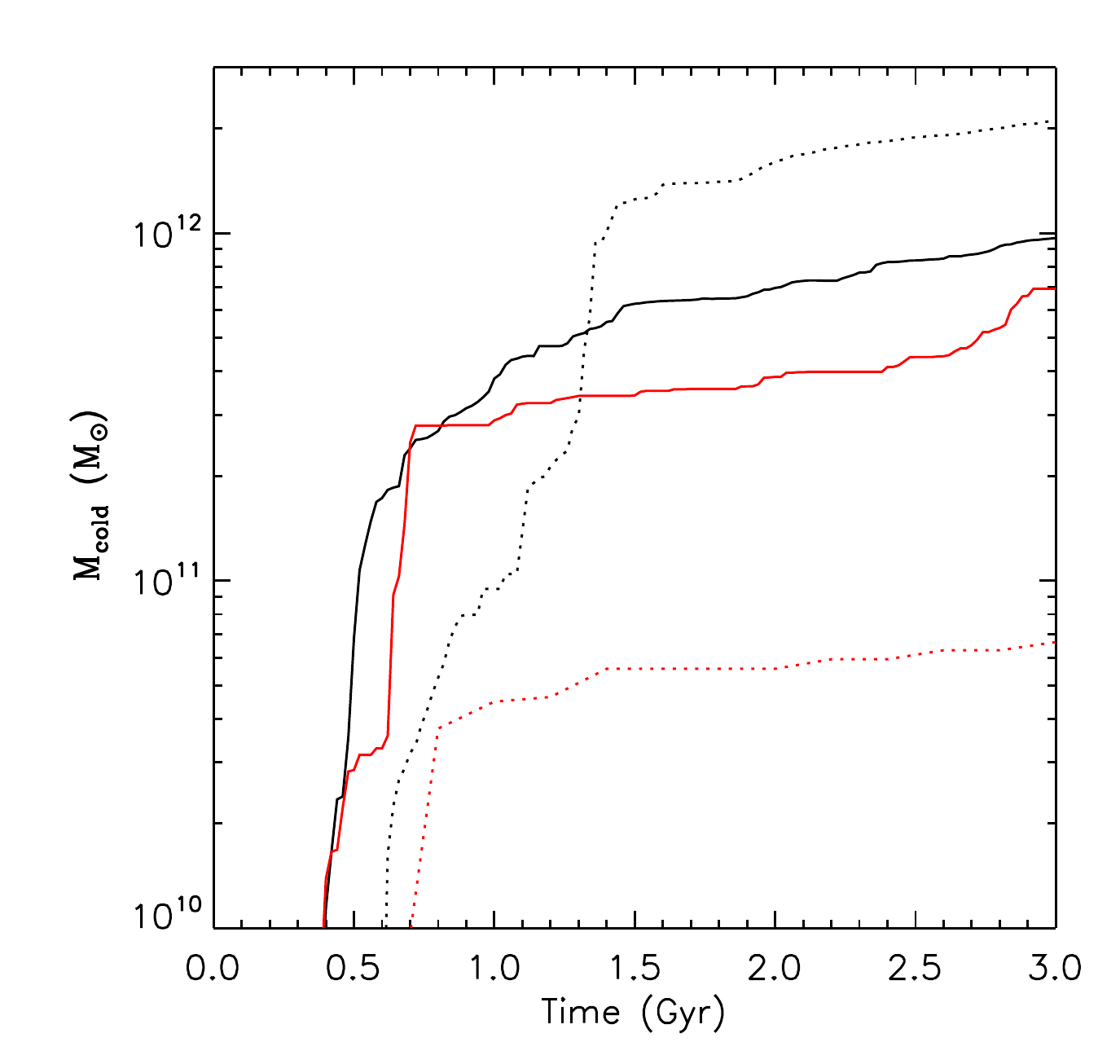} 
\caption{Total mass of cold gas as a function of time for the conductive (red) and non-conductive (black) simulations. The data for the high-mass cluster (Run CAM$h$ and Run AM$h$) and the Perseus-like cluster (Run CA and Run A) is plotted using dotted and solid lines, respectively.}
\label{fig:Mcold}
\end{center}
\end{figure}

Indeed, the amount of cold gas (i.e., the summation of masses of all cold-gas particles in our simulations) is significantly suppressed in the conductive case, as can be seen from Figure \ref{fig:Mcold}. While the amount of cold gas differs only by a factor of $\sim 1.4$ at $t=3$\ Gyr for the Perseus-like cluster, it is suppressed by a factor of $\sim 31.7$ for the high-mass cluster in the runs with conduction. 

In terms of mass dependence (for the same feedback efficient parameter $\epsilon$), simulations with and without conduction have several distinct trends. 
When there is no conduction, the AGN could more easily displace the gas, heat the ICM, and elevate the entropy profiles for the lower-mass clusters, but have to strive to balance cooling and maintain the entropy profiles near the minimum in the higher-mass systems. As a result, the higher-mass clusters spend more time in the $t_{\rm c}/t_{\rm ff} \lesssim 10$ states which allows for local TI to occur, and thus a monotonic increase in the amount of cold gas with cluster masses is found (black lines in Figure \ref{fig:Mcold}). One might expect the duty cycle to increase with cluster masses too due to the same reason, but we find that it does not have an obvious trend with cluster masses (numbers quoted above), possibly because of the compensating effect that the jets in high-mass clusters have somewhat higher thermalization efficiencies (e.g., comparing bottom panels of Figure \ref{fig:heating} and Figure \ref{fig:heating_M15}). On the contrary, cored entropy profiles are seen in the conductive clusters of all masses. Because of the larger $t_{\rm c}/t_{\rm ff}$ and suppressed TI due to conduction, the AGN duty cycle and the amount of cold gas both drop significantly as the cluster mass increases.  

These trends, especially the mass dependence of AGN duty cycles and the amount of cold gas in CC clusters, in principle could be tested against with observations to determine the significance of thermal conduction in clusters. Although direct comparisons may be difficult because of, for example, the different definitions of duty cycles in the simulations and observations, the competition between the cold gas and star formation \citep[e.g.,][]{Li15}, the exact cosmological histories of the clusters, and limitations of our numerical techniques (see \S~\ref{sec:coldgas}), our simulations suggest that if the above trends exist above the mass of the Perseus cluster, it would be a supporting evidence for unimpeded thermal conduction within clusters. 


\subsection{Parameter dependence}
\label{sec:param}

In the following we discuss the dependence of our results on the feedback efficiency parameter ($\epsilon$ in Eq.\ \ref{eq:jet}) and the accretion model (namely, the hot mode accretion). 

\begin{figure*}[tbp]
\begin{center}
\includegraphics[scale=0.55]{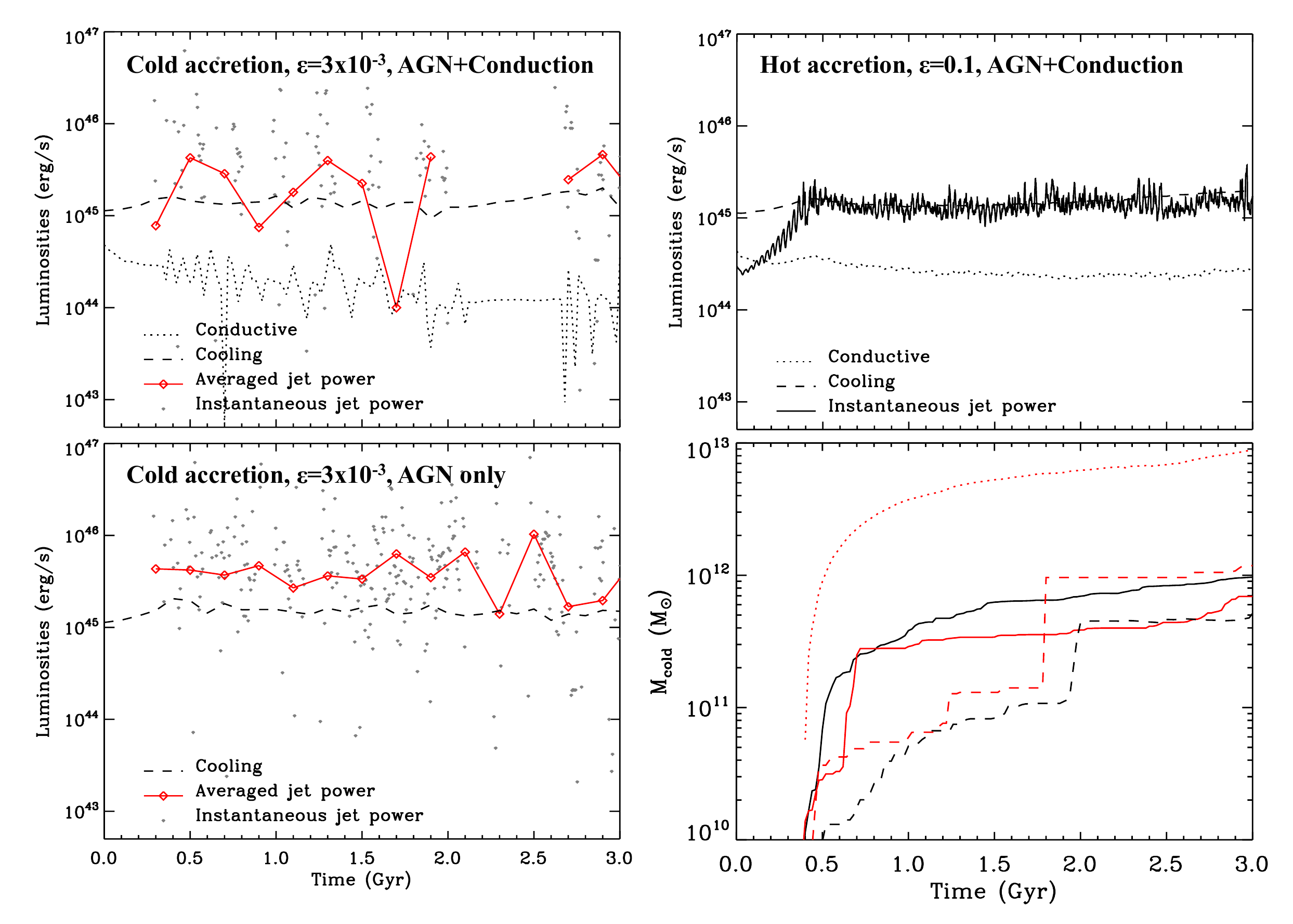} 
\caption{Dependence of results on the feedback efficiency parameter ($\epsilon$) and the accretion model. {\it Left column:} Luminosity components for the simulations with (top) and without (bottom) conduction (meaning of lines same as in Figure \ref{fig:heating}). All parameters are the same as used in Run CA and Run A expect that $\epsilon=3\times10^{-3}$. {\it Top right:} Luminosity components for the simulation adopting the hot-mode accretion. {\it Bottom right:} Total amount of cold gas in these simulations compared to the fiducial cases (solid lines). The results for $\epsilon=3\times10^{-3}$ and hot accretion are shown in dashed and dotted lines, respectively. Red curves represent simulations including conduction.}
\label{fig:param}
\end{center}
\end{figure*}

Simulations with a higher feedback efficiency $\epsilon=3\times10^{-3}$ are shown in the left panels of Figure \ref{fig:param}. Compared with the fiducial runs shown in Figure \ref{fig:heating}, we find that for both the conductive and non-conductive runs, the main difference is the frequency of AGN outbursts. Namely, in order to compensate for the increased feedback efficiency and thus larger instantaneous jet power, the AGN self-adjusts and inject energy less frequently (shown as grey dots), as also found in other simulations of cold-mode feedback \citep{Gaspari11, Li14, Prasad15}. The amount of cold gas is in general suppressed for higher $\epsilon$ (bottom right panel, dashed lines) because the temperature, entropy, and $t_{\rm c}/t_{\rm ff}$ profiles are more likely to be elevated due to strong outbursts. However, the trend is sometimes reversed because the AGN activity is highly variable. For instance, the conductive simulation with higher $\epsilon$ (red dashed line) shows a sudden increase in the total cold gas mass at $t\sim 1.8$\ Gyr due to substantial ICM cooling after a period of quiescent AGN activity.  

The results obtained using the hot mode accretion is presented in the top right panel of Figure \ref{fig:param}. Initially the cluster cools and contracts because cooling dominates over AGN plus conductive heating. The SMBH accretion rate thus increases until the jet power balances the cooling losses at $t\sim 0.5$\ Gyr. Since the feedback is continuous, it is essentially equivalent to the cold-mode accretion but with a very small feedback efficiency (so that the energy injection is almost continuous). Indeed, the temperature, entropy, and $t_{\rm c}/t_{\rm ff}$ profiles stay close to the minimum throughout the simulation, enabling cold gas formation at all times and causing the large amount of cold gas at the end of the simulation (red dotted line in the bottom right panel). Most of the cooled ICM is expected to form stars or be reheated by supernova feedback, with the remaining staying as cold gas \citep{Li15}. However, such a large mass of cooled ICM predicted by the simulation of hot accretion may still be difficult to reconcile with both the observed star formation rate \citep[e.g.][]{Hoffer12} and cold gas mass in the typical CC clusters \citep{Edge01, Salome03}. This conclusion is consistent with the finding of \cite{Li15} that weaker, more continuous AGN jets (modeled by the hot accretion or cold accretion with too small efficiencies) are unfavorable because they would predict excessive star formation and/or cold gas near the cluster centers. Instead, more powerful and episodic AGN outbursts (using a higher proportionality factor in the Bondi accretion rate as in Eq.\ \ref{eq:bondi} or a higher feedback efficiency $\epsilon$) would be required to reproduce the observed data. 

Despite the distinct AGN outburst histories, we find that our conclusions regarding the relative importance of conductive to AGN heating are robust. As found in \S~\ref{sec:heating}, conductive heating decreases with time due to the HBI and reduced temperature gradient in the simulation with higher feedback efficiency (top left panel in Figure \ref{fig:param}). For the hot-accretion model (top right panel), conductive heating does not drop as much because the AGN is always active from the beginning of the simulation, reorienting the magnetic field lines along the jet axis (similar to the bottom row in Figure \ref{fig:AGNslices}). However, the amount of conductive heating is still held at a subdominant level compared to AGN heating. It is remarkable that these results are insensitive to the uncertainties in the AGN accretion and feedback prescriptions.  

   

\section{Discussion}
\label{sec:discussion}

\subsection{Distribution of cold gas}
\label{sec:coldgas}

The main focus on this paper is on the evolution of the hot ICM in response to AGN feedback and anisotropic thermal conduction. However, there have been growing interests in the process of TI and how the resulting cold clumps feed and interact with the central SMBH, motivated by recent advancement both theoretically and observationally (see \S~\ref{sec:AGN}). We hereby discuss the distribution of cold gas found in our simulations and the implications. Note that the cold gas ($T < 5\times 10^5$\ K) in the simulations is dropped out from the hot phase and replaced by passively evolving particles. The assumptions and limitations of this approach are discussed at the end of this section. 

\begin{figure*}[tbp]
\begin{center}
\includegraphics[scale=0.8]{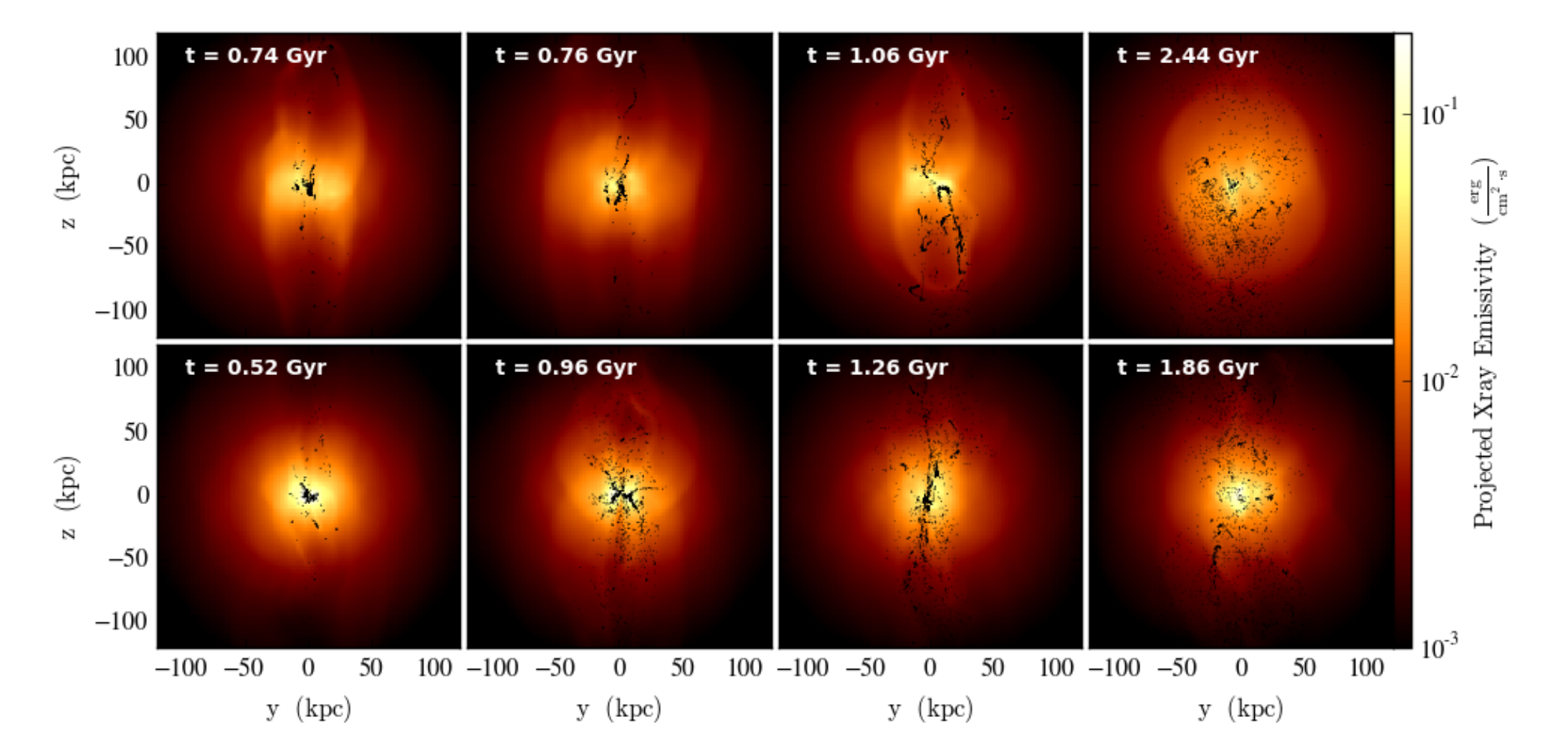} 
\caption{Snapshots of projected X-ray emissivity overplotted with projected locations of the cold gas particles for the conductive (top row, Run CA) and non-conductive (bottom row, Run A) simulations. The morphology of the cold gas is diverse, ranging from compact, filamentary, linear, to dispersed (from left to right columns, respectively), similar to observed systems. }
\label{fig:cold}
\end{center}
\end{figure*}

Figure \ref{fig:cold} shows the distribution of the cold gas particles overlaid with the projected X-ray emissivity for different epochs in the conductive (top row, Run CA) and non-conductive (bottom row, Run A) simulations of the Perseus-like cluster. The snapshots are chosen to represent the diverse morphology of the cold gas, including compact, filamentary, linear, and dispersed distributions (from left to right columns, respectively). Such a wide range of structures are also found in observations of cold gas \citep[e.g.,][]{Donahue00, McDonald11, Werner14} and stellar populations \citep[e.g.][]{Donahue15, Tremblay15} in CC clusters. In particular, the long, nearly isotropic filaments extending tens of kpc in the Perseus cluster \citep{Conselice01} are reproduced (second panel in the bottom row). 

The primary mechanisms for determining the evolution of the cold gas in our simulations are uplifting by the AGN outflows and gas motions in the ICM. When the cluster core experiences rapid cooling and meets the criterion for TI, condensation of cold gas is facilitated. A small fraction of it feeds the SMBH and triggers the AGN, while the rest can be uplifted by the jet-blown bubbles and form short-lived filamentary structures in the wakes of the bubbles \citep{Churazov01}. The cold gas is then displaced by shocks, sound waves, turbulence, and large-scale circulations in the ICM, while new filaments are formed due to subsequent AGN activity. These processes occur both in the non-conductive and conductive simulations, which is why the cold gas distributions for Run CA and Run A are not significantly different. Note also that the `filaments' seen in Figure \ref{fig:cold} are projected superpositions of cold `clouds'. The cluster core, though in quasi-equilibrium (e.g., Figure \ref{fig:heating}), is very dynamic and thus enables the complex morphology of the cold gas. Due to the above reasons, there appears to be an evolutionary trend that more compact distributions tend to be in younger clusters, while long filaments and dispersed structures preferentially occur in more evolved systems (left to right columns in Figure \ref{fig:cold}). Note that, for some observed clusters, no traces of cold gas are detected \citep{Edge01}. They could possibly be cases where the cold gas is so dispersed that the signal is below instrumental sensitivities. Note also that observations have found systems in which the cold gas sits in rotating disks \citep[e.g.,][]{Russell14}. However, they are absent in our simulations because self-gravity of the cold gas particles is neglected. Also, the cold gas near the cluster center is efficiently uplifted by AGN outflows (see discussions at the end of the section).  

There has been an outstanding debate about the formation of filaments of cold gas in CC clusters. Two contending mechanisms are uplifting by AGN outflows \citep[e.g.,][]{Churazov01, Hatch06} and in-situ condensation due to local TI \citep[e.g.,][]{Sharma10}. Our simulations suggest that the former is capable of producing extended filaments out to radius of $40-50$ kpc from the cluster center, as seen in some clusters \citep{Conselice01, McDonald11, McDonald15}. On the other hand, the cold gas is condensed out only within the inner 20-30 kpc, consistent with the radius within which $t_{\rm c}/t_{\rm ff} \lesssim 10$ ($\equiv r_{\rm TI}$). Therefore, while local TI could possibly explain filaments within $r_{\rm TI}$ \citep{McDonald11}, for filaments that extend beyond $r_{\rm TI}$ \citep{McDonald15, Tremblay15}, AGN uplifting could be essential. 

We also note that the filaments shown in Figure \ref{fig:cold} do not necessarily trail behind cavities in the images (though they originally did and some still do), as one would expect from a classical picture of AGN uplifting \citep{Churazov01}. Instead, the filaments could be isotropic (e.g., second panel in the bottom row), off-axis (e.g., bottom row, third panel), or on the side of the cavities (e.g., top row, third panel), again attributed to repeated AGN activity and the complex gas motions within the cluster core. For example, the third panel in the top row in Figure \ref{fig:cold} shows a system similar to A1795 \citep{McDonald09}, with a prominent filament reaching to the SW direction (the observed filament extends to the SE direction and lies next to a `hook' structure that might be associated with an AGN-blown bubble). In the simulations we find that the cold gas within that particular filament was ejected by a previous event of the AGN. While it then falls back a little during a short, quiescent phase of AGN activity, it is pushed to the side of the south cavity due to shocks generated by the latest outburst. Therefore, AGN uplifting could still be responsible for the formation of the filaments even though they are not found in the wakes of the current cavities \citep[e.g.,][]{McDonald15}. 

There are of course several caveats since our simulations do not directly simulate the cold gas but only treat them as passively evolving particles. First, the detailed process of uplifting the cold gas by AGN outflows is not accounted for self-consistently. Effectively, our approach probes an extreme case where AGN uplifting is very efficient, perhaps due to a strong drag force by the hot ICM, or if the size of the cold clumps is very small \citep[e.g.,][who have shown that relativistic jets can efficiently disperse cold clumps with sizes $\lesssim 50$\ pc]{Wagner12}. The latter may be justified by that fact that, in a simulation with infinite resolution, the clump sizes should be of the order of the Field length in the cold medium ($\sim 10^4$\ K), which is $\sim 10^{-4}$\ pc \citep{Sharma10}. Observationally, there is also much evidence for cold gas being uplifted by buoyantly rising bubbles \citep{Fabian03, Hatch06, Canning11, Werner13}, suggesting that the cold gas does not simply feel the gravitational force, but has a response to the hydrodynamic forces from the hot gas. Second, there are no destruction mechanisms within our simulated clusters once the temperature of the cold gas drops below $T=5\times10^5$\ K. Therefore, the cold gas cannot be destroyed by shocks \citep[][]{Klein94, Stone92, Cooper09} or evaporated by thermal conduction \citep[e.g.,][]{Orlando05, Orlando08}, and thus the amount of cold gas obtained in the simulations (e.g., Figure \ref{fig:Mcold}) likely overestimates the actual value. Finally, our simulations are unable to follow the exact process of TI down to the Field length of the cold phase, which is far beyond the resolution of our simulations. Consequently, the detailed statistics of the cold clumps formed in the simulations may not be converged \citep[e.g.,][]{Koyama04}. This is also why our conductive simulations do not show an obvious evidence for filaments forming because of the suppression of TI along the magnetic field directions \citep{Sharma10, Wagh14}. To accurately simulate the formation of filaments from TI would require extremely high-resolution simulations of TI that account for cooling, anisotropic conduction, pressure support from magnetic fields and cosmic rays, and effects of external turbulence, which is beyond the scope of this paper. We thus limit the focus on the paper on the hot ICM and only derive qualitative trends when discussing the cold phase.    


\subsection{Comparisons with previous works}
\label{sec:comparison}

Earlier simulations that investigated the effects of HBI \citep{Parrish09, Bogdanovic09, Ruszkowski10, McCourt11, Kunz12, Avara13} have focused on scenarios where the magnetic field is negligible ($\beta > 10^5$). We use a realistic magnetic field strength ($\beta \sim 100$) for typical ICM and show that magnetic tension could suppress a significant portion of the HBI unstable modes below a critical length scale (Eq.\ \ref{eq:lcrit}). Furthermore, whether the HBI could develop at all would depend on the actual scale of magnetic field perturbations (i.e., the unknown magnetic field coherence length $l_{\rm B}$). Though in the present paper we do not include the effect of anisotropic viscosity, \cite{Kunz12} has shown that it suppresses the HBI in a similar way as magnetic tension. In either case, the field-line wrapping effect of the HBI should be substantially impaired for realistic ICM conditions. 

If the HBI is still allowed to grow (e.g., $l_{\rm B}>l_{\rm crit}$ or when the ICM is collisional so that effects of anisotropic viscosity is attenuated), \cite{Ruszkowski10} showed that the magnetic field lines could be effectively randomized if the turbulence in the ICM is volume-filling. Without an initially imposed turbulent velocity field, we showed that the AGN jet-driven turbulence is able to stir up the field lines and sustain conductivity at 1/3 of Spitzer. However, the turbulence developed is contained only in the cones along the jet axis and is not volume-filling, because the frequency distribution of AGN outbursts does not allow efficient excitation of the $g$-modes outside the cones of influence of the AGN. 

Our work is the first MHD simulations of cluster CCs that include radiative cooling, AGN feedback, and anisotropic thermal conduction. We hereby compare the results from our non-conductive simulations to previous works of self-regulated AGN feedback in CC clusters. Our prescriptions of SMBH accretion largely inherit from previous simulations, including both the hot \citep[e.g.,][]{Sijacki07, Cattaneo07, Dubois10, Yang12b} and cold \citep[e.g.,][]{Gaspari11, Li14, Prasad15} modes. The AGN feedback is purely kinetic and modeled as bipolar jets, which allows the preservation of the CC structure \citep{Gaspari11}. 

The implementation of our cold-mode feedback and initial conditions are closest to the work by \cite{Li14}, though there are still several distinctions. First, our simulations include magnetic fields while theirs are purely hydrodynamic. As a result, the AGN bubbles in our simulations in general keep their integrity for longer periods of time, reach to further distances, and mix with the ambient ICM more slowly, because of the suppression of hydrodynamic instabilities by magnetic tension at the surface of the bubbles \citep[e.g.,][]{Robinson04}. Because it is more difficult for the bubbles to couple with the ICM, in general a lower feedback efficiency $\epsilon$ (thus a smaller radius of influence and higher thermalization efficiency) needs to be used in the MHD simulations in order to generate similar profiles as in hydrodynamic simulations. Moreover, the initial tangled magnetic field inserts a small random velocity field that facilitates local TI \citep{Singh15}, which has a similar effect as introducing initial density perturbations \citep[e.g.,][]{Gaspari12, Prasad15}. Therefore, our simulations show cold clump formation in all directions, rather than confined with the jet cones \citep{Li14}. Another benefit of starting with nonzero initial perturbations is that the early evolution of the ICM is less sensitive to the feedback efficiency parameter $\epsilon$, while in the absence of initial perturbations, the feedback efficiency has to be large enough for the first jet events to stir up the medium enough to permit further clump formation. Lastly, the cold gas in our simulations is dropped out from the hot ICM \citep{Gaspari11}, while other recent studies \citep{Li14, Li15, Prasad15} have focused on TI-regulated feedback and simulated the cold medium directly. This causes a major difference in the ICM evolution. Because our approach essentially assumes efficient AGN uplifting of the cold gas (see the end of \S~\ref{sec:coldgas}), the cold gas is more easily entrained in the bubble wakes and distributed within the cluster cores. In contrast, the other simulations commonly found rotating disks of cold gas accumulated near the cluster center, which could feed the SMBH and induce more variable AGN outburst histories. Since none of the existing simulations are able to resolve the size of the cold clumps ($\sim 10^{-4}$\ pc, which is the Field length of the cold gas), the efficiency of AGN uplifting and thus the long-term evolution of the cold disks and the ICM are still uncertain. We note, though, that our main conclusions do not depend on the details of AGN accretion and feedback prescriptions (\S~\ref{sec:param}). Additionally, the more efficient AGN uplifting of cold gas allows more direct coupling between cooling from the ICM and AGN feedback, quickly establishing a quasi-equilibrium state that is optimal for the purpose of our study.


\section{Conclusions}
\label{sec:conclusion}

Cool cores of galaxy clusters involve complex astrophysical processes, including radiative cooling, AGN feedback, and their interactions with the magnetized ICM. In order to understand the interplay among cooling, AGN feedback, and anisotropic thermal conduction, we performed the first 3D MHD simulations including these processes in idealized cluster atmospheres. Our main findings are summarized as follows. 

1.\ For realistic magnetic field strengths in clusters ($\beta \sim 100$), magnetic tension is able to suppress the HBI for scales under a critical length ($l_{\rm crit}$, see Eq.\ \ref{eq:lcrit}). The regions stable to the HBI can form magnetic `streams' (as opposed to magnetic `filaments' seen in previous high $\beta$ simulations) and sustain conductivity with an effective Spitzer fraction of $\gtrsim 0.2$. In addition, the HBI would not develop at all if the (unknown) cluster magnetic fields are tangled on scales below $l_{\rm crit}$. 

2.\ Turbulence generated by repeated AGN jets can effectively randomize magnetic field lines and permit heat fluxes at $\sim 1/3$ of the Spitzer value, no matter whether the AGN is active or quiescent (see Figure \ref{fig:thetaB_AGN}). However, the AGN-driven turbulence is not volume-filling and affects only regions directly influenced by the jets (Figure \ref{fig:AGNslices}). 

3.\ For Perseus-like clusters, conductive heating is $\sim 10\%$ of the radiative losses within the cluster cores, while the remaining is balanced by AGN heating. Owing to the strong temperature dependence of thermal conduction, conducive heating can contribute to $\sim 50\%$ of the radiative losses for clusters doubling the mass of the Perseus. In general, we find that thermal conduction tends to limit itself by reducing the temperature gradients within the cores, which plays a more important role in terms of long-term evolution of the clusters than the temporary enhancement of Spitzer fraction enabled by AGN-driven turbulence.   

4.\ For the hottest clusters in which conductive heating is significant, the AGN activity and ICM properties can be greatly influenced. The effects of conduction include weaker averaged AGN jet power relative to the core X-ray luminosity, reduction of AGN duty cycles, more elevated temperature and entropy ICM profiles (Figure \ref{fig:M15prf}), and suppressed cold gas formation (Figure \ref{fig:Mcold}). Any observational evidence for these trends beyond the mass of Perseus could indicate unimpeded thermal conduction in clusters.   

5.\ For simulations that adopt cold gas-regulated AGN feedback, the distribution of cold gas exhibits a diverse morphology as seen in observations (Figure \ref{fig:cold}), including systems similar to the extended, nearly isotropic optical filaments observed in Perseus. In our simulations, we find that the primary mechanisms for forming such a variety of structures are AGN uplifting and gas motions in the ICM (see \S~\ref{sec:coldgas}).

The combined results from our simulations and previous works suggest that the HBI is either completely inhibited (if cluster magnetic fields have small coherence lengths) or significantly impaired (due to magnetic tension, the effects of anisotropic viscosity \citep{Kunz12}, and turbulence driven by AGN jets (this work), galaxy motions \citep{Ruszkowski10}, or decay from large-scale turbulence from cluster mergers \citep{Heinz10, Vazza13}). Therefore, the magnetic field lines in cluster cores are likely randomized instead of wrapped azimuthally by the HBI, yielding an effective Spitzer fraction of $\sim 1/3$ (assuming unhindered Spitzer conductivity along field lines). 

In terms of conductive heating, we find that due to the canceling effects of enhanced Spitzer fraction by AGN-driven turbulence and reduced temperature gradients, the amount of conductive heating over long terms tends to decrease with time or at best remain at its initial level. Therefore, the relative importance of conductive to AGN heating is largely determined by the temperature profile and the mass of the clusters. Previous works have tried to evaluate the importance of conductive heating from the observed profiles of clusters assuming 1/3 of Spitzer conductivity \citep[e.g.,][]{Voigt04, Voit11}. Our results suggest that there indeed exists a quasi-steady thermal balance among cooling, AGN feedback and conduction \citep{Voit11}, and the contributions from conduction and AGN heating derived from such observations should not vary more than a factor of a few over the course of the cluster evolution once their cores have reached such a quasi-equilibrium stage.    

In this paper we have assumed the most optimistic case of full Spitzer conductivity along the magnetic field lines and investigated its possible impacts (point 4 above). Future observations could provide tests of this scenario and put constraints on the true level of conductivity. We have neglected the effects of anisotropic viscosity, which is important for the outer part of the cluster cores where the ICM is weakly collisional. How it would modify the picture of AGN feedback requires future investigation. Finally, our work suggests that AGN heating is still the primary mechanism for counteracting radiative cooling for most clusters (except the hottest ones). However, how exactly the feedback energy is converted into heat (e.g., cavity heating, shock heating, sound wave dissipation, turbulent mixing, turbulent heating, or cosmic ray heating) remains uncertain. Some of these questions will be addressed in our upcoming work. 



\acknowledgments

We thank an anonymous referee for useful comments that helped improve the paper. The authors would like to thank Paul Nulsen, Brian McNamara, Matt Kunz, Yuan Li, Brian Morsony and Mark Avara for helpful discussion. HYKY acknowledges support by NASA through Einstein Postdoctoral Fellowship grant number PF4-150129 awarded by the Chandra X-ray Center, which is operated by the Smithsonian Astrophysical Observatory for NASA under contract NAS8-03060. CSR is grateful for financial support from the Simons Foundation (through a Simons Fellowship in Theoretical Physics) and the National Science Foundation (under grant AST1333514).  The simulations presented in this paper were performed on the {\tt Deepthought2} cluster, maintained and supported by the Division of Information Technology at the University of Maryland College Park. FLASH was developed largely by the DOE-supported ASC/Alliances Center for Astrophysical Thermonuclear Flashes at the University of Chicago.  


\bibliography{agn}

\begin{thebibliography}{108}
\expandafter\ifx\csname natexlab\endcsname\relax\def\natexlab#1{#1}\fi

\bibitem[{{Allen} {et~al.}(2006){Allen}, {Dunn}, {Fabian}, {Taylor}, \&
  {Reynolds}}]{Allen06}
{Allen}, S.~W., {Dunn}, R.~J.~H., {Fabian}, A.~C., {Taylor}, G.~B., \&
  {Reynolds}, C.~S. 2006, \mnras, 372, 21

\bibitem[{{Avara} {et~al.}(2013){Avara}, {Reynolds}, \&
  {Bogdanovi{\'c}}}]{Avara13}
{Avara}, M.~J., {Reynolds}, C.~S., \& {Bogdanovi{\'c}}, T. 2013, \apj, 773, 171

\bibitem[{{B{\i}rzan} {et~al.}(2013){B{\i}rzan}, {Rafferty}, {Nulsen},
  {et~al.}}]{Birzan13}
{B{\i}rzan}, L., {Rafferty}, D.~A., {Nulsen}, P.~E.~J., {et~al.} 2013,
  Astronomische Nachrichten, 334, 390

\bibitem[{{Bogdanovi{\'c}} {et~al.}(2009){Bogdanovi{\'c}}, {Reynolds},
  {Balbus}, \& {Parrish}}]{Bogdanovic09}
{Bogdanovi{\'c}}, T., {Reynolds}, C.~S., {Balbus}, S.~A., \& {Parrish}, I.~J.
  2009, \apj, 704, 211

\bibitem[{{Bregman} \& {David}(1988)}]{Bregman88}
{Bregman}, J.~N., \& {David}, L.~P. 1988, \apj, 326, 639

\bibitem[{{Canning} {et~al.}(2011){Canning}, {Fabian}, {Johnstone}, {Sanders},
  {Crawford}, {Ferland}, \& {Hatch}}]{Canning11}
{Canning}, R.~E.~A., {Fabian}, A.~C., {Johnstone}, R.~M., {Sanders}, J.~S.,
  {Crawford}, C.~S., {Ferland}, G.~J., \& {Hatch}, N.~A. 2011, \mnras, 417,
  3080

\bibitem[{{Cardiel} {et~al.}(1998){Cardiel}, {Gorgas}, \&
  {Aragon-Salamanca}}]{Cardiel98}
{Cardiel}, N., {Gorgas}, J., \& {Aragon-Salamanca}, A. 1998, \mnras, 298, 977

\bibitem[{{Carilli} \& {Taylor}(2002)}]{Carilli02}
{Carilli}, C.~L., \& {Taylor}, G.~B. 2002, \araa, 40, 319

\bibitem[{{Cattaneo} \& {Teyssier}(2007)}]{Cattaneo07}
{Cattaneo}, A., \& {Teyssier}, R. 2007, \mnras, 376, 1547

\bibitem[{{Cavagnolo} {et~al.}(2009){Cavagnolo}, {Donahue}, {Voit}, \&
  {Sun}}]{Cavagnolo}
{Cavagnolo}, K.~W., {Donahue}, M., {Voit}, G.~M., \& {Sun}, M. 2009, \apjs,
  182, 12

\bibitem[{{Chandran} \& {Cowley}(1998)}]{Chandran98}
{Chandran}, B.~D.~G., \& {Cowley}, S.~C. 1998, Physical Review Letters, 80,
  3077

\bibitem[{{Churazov} {et~al.}(2001){Churazov}, {Br{\"u}ggen}, {Kaiser},
  {B{\"o}hringer}, \& {Forman}}]{Churazov01}
{Churazov}, E., {Br{\"u}ggen}, M., {Kaiser}, C.~R., {B{\"o}hringer}, H., \&
  {Forman}, W. 2001, \apj, 554, 261

\bibitem[{{Churazov} {et~al.}(2003){Churazov}, {Forman}, {Jones}, \&
  {B{\"o}hringer}}]{Churazov03}
{Churazov}, E., {Forman}, W., {Jones}, C., \& {B{\"o}hringer}, H. 2003, \apj,
  590, 225

\bibitem[{{Conselice} {et~al.}(2001){Conselice}, {Gallagher}, \&
  {Wyse}}]{Conselice01}
{Conselice}, C.~J., {Gallagher}, III, J.~S., \& {Wyse}, R.~F.~G. 2001, \aj,
  122, 2281

\bibitem[{{Cooper} {et~al.}(2009){Cooper}, {Bicknell}, {Sutherland}, \&
  {Bland-Hawthorn}}]{Cooper09}
{Cooper}, J.~L., {Bicknell}, G.~V., {Sutherland}, R.~S., \& {Bland-Hawthorn},
  J. 2009, \apj, 703, 330

\bibitem[{{Crawford} {et~al.}(1999){Crawford}, {Allen}, {Ebeling}, {Edge}, \&
  {Fabian}}]{Crawford99}
{Crawford}, C.~S., {Allen}, S.~W., {Ebeling}, H., {Edge}, A.~C., \& {Fabian},
  A.~C. 1999, \mnras, 306, 857

\bibitem[{{Donahue} {et~al.}(2015){Donahue}, {Connor}, {Fogarty},
  {et~al.}}]{Donahue15}
{Donahue}, M., {Connor}, T., {Fogarty}, K., {et~al.} 2015, \apj, 805, 177

\bibitem[{{Donahue} {et~al.}(2000){Donahue}, {Mack}, {Voit}, {Sparks},
  {Elston}, \& {Maloney}}]{Donahue00}
{Donahue}, M., {Mack}, J., {Voit}, G.~M., {Sparks}, W., {Elston}, R., \&
  {Maloney}, P.~R. 2000, \apj, 545, 670

\bibitem[{{Dubey} {et~al.}(2008){Dubey}, {Reid}, \& {Fisher}}]{Dubey08}
{Dubey}, A., {Reid}, L.~B., \& {Fisher}, R. 2008, Physica Scripta, T132, 014046

\bibitem[{{Dubois} {et~al.}(2010){Dubois}, {Devriendt}, {Slyz}, \&
  {Teyssier}}]{Dubois10}
{Dubois}, Y., {Devriendt}, J., {Slyz}, A., \& {Teyssier}, R. 2010, \mnras, 409,
  985

\bibitem[{{Dursi} \& {Pfrommer}(2008)}]{Dursi08}
{Dursi}, L.~J., \& {Pfrommer}, C. 2008, \apj, 677, 993

\bibitem[{{Edge}(2001)}]{Edge01}
{Edge}, A.~C. 2001, \mnras, 328, 762

\bibitem[{{Fabian}(1994)}]{Fabian94}
{Fabian}, A.~C. 1994, ARA\&A, 32, 277

\bibitem[{{Fabian} {et~al.}(2003){Fabian}, {Sanders}, {Crawford}, {Conselice},
  {Gallagher}, \& {Wyse}}]{Fabian03}
{Fabian}, A.~C., {Sanders}, J.~S., {Crawford}, C.~S., {Conselice}, C.~J.,
  {Gallagher}, J.~S., \& {Wyse}, R.~F.~G. 2003, \mnras, 344, L48

\bibitem[{{Fryxell} {et~al.}(2000){Fryxell}, {Olson}, {Ricker},
  {et~al.}}]{Flash}
{Fryxell}, B., {Olson}, K., {Ricker}, P., {et~al.} 2000, \apjs, 131, 273

\bibitem[{{Gaspari} {et~al.}(2015){Gaspari}, {Brighenti}, \&
  {Temi}}]{Gaspari15}
{Gaspari}, M., {Brighenti}, F., \& {Temi}, P. 2015, \aap, 579, A62

\bibitem[{{Gaspari} {et~al.}(2011){Gaspari}, {Melioli}, {Brighenti}, \&
  {D'Ercole}}]{Gaspari11}
{Gaspari}, M., {Melioli}, C., {Brighenti}, F., \& {D'Ercole}, A. 2011, \mnras,
  411, 349

\bibitem[{{Gaspari} {et~al.}(2013){Gaspari}, {Ruszkowski}, \& {Oh}}]{Gaspari13}
{Gaspari}, M., {Ruszkowski}, M., \& {Oh}, S.~P. 2013, \mnras, 432, 3401

\bibitem[{{Gaspari} {et~al.}(2012){Gaspari}, {Ruszkowski}, \&
  {Sharma}}]{Gaspari12}
{Gaspari}, M., {Ruszkowski}, M., \& {Sharma}, P. 2012, \apj, 746, 94

\bibitem[{{Hamer} {et~al.}(2014){Hamer}, {Edge}, {Swinbank},
  {et~al.}}]{Hamer14}
{Hamer}, S.~L., {Edge}, A.~C., {Swinbank}, A.~M., {et~al.} 2014, \mnras, 437,
  862

\bibitem[{{Hatch} {et~al.}(2006){Hatch}, {Crawford}, {Johnstone}, \&
  {Fabian}}]{Hatch06}
{Hatch}, N.~A., {Crawford}, C.~S., {Johnstone}, R.~M., \& {Fabian}, A.~C. 2006,
  \mnras, 367, 433

\bibitem[{{Heinz} {et~al.}(2010){Heinz}, {Br{\"u}ggen}, \& {Morsony}}]{Heinz10}
{Heinz}, S., {Br{\"u}ggen}, M., \& {Morsony}, B. 2010, \apj, 708, 462

\bibitem[{{Hobbs} {et~al.}(2012){Hobbs}, {Power}, {Nayakshin}, \&
  {King}}]{Hobbs12}
{Hobbs}, A., {Power}, C., {Nayakshin}, S., \& {King}, A.~R. 2012, \mnras, 421,
  3443

\bibitem[{{Hoffer} {et~al.}(2012){Hoffer}, {Donahue}, {Hicks}, \&
  {Barthelemy}}]{Hoffer12}
{Hoffer}, A.~S., {Donahue}, M., {Hicks}, A., \& {Barthelemy}, R.~S. 2012,
  \apjs, 199, 23

\bibitem[{{Kaiser}(1986)}]{Kaiser86}
{Kaiser}, N. 1986, \mnras, 222, 323

\bibitem[{{Klein} {et~al.}(1994){Klein}, {McKee}, \& {Colella}}]{Klein94}
{Klein}, R.~I., {McKee}, C.~F., \& {Colella}, P. 1994, \apj, 420, 213

\bibitem[{{Koyama} \& {Inutsuka}(2004)}]{Koyama04}
{Koyama}, H., \& {Inutsuka}, S.-i. 2004, \apjl, 602, L25

\bibitem[{{Kuchar} \& {En{\ss}lin}(2011)}]{Kuchar11}
{Kuchar}, P., \& {En{\ss}lin}, T.~A. 2011, \aap, 529, A13

\bibitem[{{Kunz}(2011)}]{Kunz11}
{Kunz}, M.~W. 2011, \mnras, 417, 602

\bibitem[{{Kunz} {et~al.}(2012){Kunz}, {Bogdanovi{\'c}}, {Reynolds}, \&
  {Stone}}]{Kunz12}
{Kunz}, M.~W., {Bogdanovi{\'c}}, T., {Reynolds}, C.~S., \& {Stone}, J.~M. 2012,
  \apj, 754, 122

\bibitem[{{Kunz} {et~al.}(2014){Kunz}, {Schekochihin}, \& {Stone}}]{Kunz14}
{Kunz}, M.~W., {Schekochihin}, A.~A., \& {Stone}, J.~M. 2014, Physical Review
  Letters, 112, 205003

\bibitem[{{Lee}(2013)}]{Lee13}
{Lee}, D. 2013, Journal of Computational Physics, 243, 269

\bibitem[{Lee \& Deane(2009)}]{Lee09}
Lee, D., \& Deane, A.~E. 2009, Journal of Computational Physics, 228, 952

\bibitem[{{Li} \& {Bryan}(2014{\natexlab{a}})}]{Li14a}
{Li}, Y., \& {Bryan}, G.~L. 2014{\natexlab{a}}, \apj, 789, 54

\bibitem[{{Li} \& {Bryan}(2014{\natexlab{b}})}]{Li14}
---. 2014{\natexlab{b}}, \apj, 789, 153

\bibitem[{{Li} {et~al.}(2015){Li}, {Bryan}, {Ruszkowski}, {et~al.}}]{Li15}
{Li}, Y., {Bryan}, G.~L., {Ruszkowski}, M., {et~al.} 2015, ArXiv: 1503.02660

\bibitem[{{Lyutikov}(2006)}]{Lyutikov06}
{Lyutikov}, M. 2006, \mnras, 373, 73

\bibitem[{{McCourt} {et~al.}(2011){McCourt}, {Parrish}, {Sharma}, \&
  {Quataert}}]{McCourt11}
{McCourt}, M., {Parrish}, I.~J., {Sharma}, P., \& {Quataert}, E. 2011, \mnras,
  413, 1295

\bibitem[{{McCourt} {et~al.}(2012){McCourt}, {Sharma}, {Quataert}, \&
  {Parrish}}]{McCourt12}
{McCourt}, M., {Sharma}, P., {Quataert}, E., \& {Parrish}, I.~J. 2012, \mnras,
  419, 3319

\bibitem[{{McDonald} {et~al.}(2015){McDonald}, {McNamara}, {van Weeren},
  {et~al.}}]{McDonald15}
{McDonald}, M., {McNamara}, B.~R., {van Weeren}, R.~J., {et~al.} 2015, ArXiv:
  1508.05941

\bibitem[{{McDonald} \& {Veilleux}(2009)}]{McDonald09}
{McDonald}, M., \& {Veilleux}, S. 2009, \apjl, 703, L172

\bibitem[{{McDonald} {et~al.}(2011){McDonald}, {Veilleux}, \&
  {Mushotzky}}]{McDonald11}
{McDonald}, M., {Veilleux}, S., \& {Mushotzky}, R. 2011, \apj, 731, 33

\bibitem[{{McNamara} \& {Nulsen}(2007)}]{McNamara07}
{McNamara}, B.~R., \& {Nulsen}, P.~E.~J. 2007, \araa, 45, 117

\bibitem[{{McNamara} {et~al.}(2011){McNamara}, {Rohanizadegan}, \&
  {Nulsen}}]{McNamara11}
{McNamara}, B.~R., {Rohanizadegan}, M., \& {Nulsen}, P.~E.~J. 2011, \apj, 727,
  39

\bibitem[{{Meece} {et~al.}(2015){Meece}, {O'Shea}, \& {Voit}}]{Meece15}
{Meece}, G.~R., {O'Shea}, B.~W., \& {Voit}, G.~M. 2015, \apj, 808, 43

\bibitem[{{Meneghetti} {et~al.}(2014){Meneghetti}, {Rasia}, {Vega},
  {et~al.}}]{Meneghetti14}
{Meneghetti}, M., {Rasia}, E., {Vega}, J., {et~al.} 2014, \apj, 797, 34

\bibitem[{{Narayan} \& {Medvedev}(2001)}]{Narayan01}
{Narayan}, R., \& {Medvedev}, M.~V. 2001, \apjl, 562, L129

\bibitem[{{Navarro} {et~al.}(1996){Navarro}, {Frenk}, \& {White}}]{Navarro96}
{Navarro}, J.~F., {Frenk}, C.~S., \& {White}, S.~D.~M. 1996, \apj, 462, 563

\bibitem[{{Orlando} {et~al.}(2008){Orlando}, {Bocchino}, {Reale}, {Peres}, \&
  {Pagano}}]{Orlando08}
{Orlando}, S., {Bocchino}, F., {Reale}, F., {Peres}, G., \& {Pagano}, P. 2008,
  \apj, 678, 274

\bibitem[{{Orlando} {et~al.}(2005){Orlando}, {Peres}, {Reale},
  {et~al.}}]{Orlando05}
{Orlando}, S., {Peres}, G., {Reale}, F., {et~al.} 2005, \aap, 444, 505

\bibitem[{{Panagoulia} {et~al.}(2014){Panagoulia}, {Fabian}, \&
  {Sanders}}]{Panagoulia14}
{Panagoulia}, E.~K., {Fabian}, A.~C., \& {Sanders}, J.~S. 2014, \mnras, 438,
  2341

\bibitem[{{Parrish} {et~al.}(2009){Parrish}, {Quataert}, \&
  {Sharma}}]{Parrish09}
{Parrish}, I.~J., {Quataert}, E., \& {Sharma}, P. 2009, \apj, 703, 96

\bibitem[{{Peterson} {et~al.}(2003){Peterson}, {Kahn}, {Paerels},
  {et~al.}}]{Peterson03}
{Peterson}, J.~R., {Kahn}, S.~M., {Paerels}, F.~B.~S., {et~al.} 2003, \apj,
  590, 207

\bibitem[{{Pinzke} \& {Pfrommer}(2010)}]{Pinzke10}
{Pinzke}, A., \& {Pfrommer}, C. 2010, \mnras, 409, 449

\bibitem[{{Pizzolato} \& {Soker}(2005)}]{Pizzolato05}
{Pizzolato}, F., \& {Soker}, N. 2005, \apj, 632, 821

\bibitem[{{Pizzolato} \& {Soker}(2010)}]{Pizzolato10}
---. 2010, \mnras, 408, 961

\bibitem[{{Pope} {et~al.}(2006){Pope}, {Pavlovski}, {Kaiser}, \&
  {Fangohr}}]{Pope06}
{Pope}, E.~C.~D., {Pavlovski}, G., {Kaiser}, C.~R., \& {Fangohr}, H. 2006,
  \mnras, 367, 1121

\bibitem[{{Prasad} {et~al.}(2015){Prasad}, {Sharma}, \& {Babul}}]{Prasad15}
{Prasad}, D., {Sharma}, P., \& {Babul}, A. 2015, ArXiv: 1504.02215

\bibitem[{{Quataert}(2008)}]{Quataert08}
{Quataert}, E. 2008, \apj, 673, 758

\bibitem[{{Rafferty} {et~al.}(2006){Rafferty}, {McNamara}, {Nulsen}, \&
  {Wise}}]{Rafferty06}
{Rafferty}, D.~A., {McNamara}, B.~R., {Nulsen}, P.~E.~J., \& {Wise}, M.~W.
  2006, \apj, 652, 216

\bibitem[{{Reynolds} {et~al.}(2015){Reynolds}, {Balbus}, \&
  {Schekochihin}}]{Reynolds15}
{Reynolds}, C.~S., {Balbus}, S.~A., \& {Schekochihin}, A.~A. 2015, \apj, 815,
  41

\bibitem[{{Robinson} {et~al.}(2004){Robinson}, {Dursi}, {Ricker},
  {et~al.}}]{Robinson04}
{Robinson}, K., {Dursi}, L.~J., {Ricker}, P.~M., {et~al.} 2004, \apj, 601, 621

\bibitem[{{Russell} {et~al.}(2014){Russell}, {McNamara}, {Edge},
  {et~al.}}]{Russell14}
{Russell}, H.~R., {McNamara}, B.~R., {Edge}, A.~C., {et~al.} 2014, \apj, 784,
  78

\bibitem[{{Ruszkowski} {et~al.}(2011){Ruszkowski}, {Lee}, {Br{\"u}ggen},
  {Parrish}, \& {Oh}}]{Ruszkowski11b}
{Ruszkowski}, M., {Lee}, D., {Br{\"u}ggen}, M., {Parrish}, I., \& {Oh}, S.~P.
  2011, \apj, 740, 81

\bibitem[{{Ruszkowski} \& {Oh}(2010)}]{Ruszkowski10}
{Ruszkowski}, M., \& {Oh}, S.~P. 2010, \apj, 713, 1332

\bibitem[{{Ruszkowski} \& {Oh}(2011)}]{Ruszkowski11a}
---. 2011, \mnras, 414, 1493

\bibitem[{{Salom{\'e}} \& {Combes}(2003)}]{Salome03}
{Salom{\'e}}, P., \& {Combes}, F. 2003, \aap, 412, 657

\bibitem[{{Salom{\'e}} {et~al.}(2006){Salom{\'e}}, {Combes}, {Edge},
  {et~al.}}]{Salome06}
{Salom{\'e}}, P., {Combes}, F., {Edge}, A.~C., {et~al.} 2006, \aap, 454, 437

\bibitem[{{Sanders} {et~al.}(2008){Sanders}, {Fabian}, {Allen},
  {et~al.}}]{Sanders08}
{Sanders}, J.~S., {Fabian}, A.~C., {Allen}, S.~W., {et~al.} 2008, \mnras, 385,
  1186

\bibitem[{{Scannapieco} \& {Br{\"u}ggen}(2008)}]{Scannapieco08}
{Scannapieco}, E., \& {Br{\"u}ggen}, M. 2008, \apj, 686, 927

\bibitem[{{Schekochihin} {et~al.}(2010){Schekochihin}, {Cowley}, {Rincon}, \&
  {Rosin}}]{Schekochihin10}
{Schekochihin}, A.~A., {Cowley}, S.~C., {Rincon}, F., \& {Rosin}, M.~S. 2010,
  \mnras, 405, 291

\bibitem[{{Sharma} \& {Hammett}(2007)}]{Sharma07}
{Sharma}, P., \& {Hammett}, G.~W. 2007, Journal of Computational Physics, 227,
  123

\bibitem[{{Sharma} {et~al.}(2012){Sharma}, {McCourt}, {Quataert}, \&
  {Parrish}}]{Sharma12}
{Sharma}, P., {McCourt}, M., {Quataert}, E., \& {Parrish}, I.~J. 2012, \mnras,
  420, 3174

\bibitem[{{Sharma} {et~al.}(2010){Sharma}, {Parrish}, \& {Quataert}}]{Sharma10}
{Sharma}, P., {Parrish}, I.~J., \& {Quataert}, E. 2010, \apj, 720, 652

\bibitem[{{Sijacki} {et~al.}(2007){Sijacki}, {Springel}, {Di Matteo}, \&
  {Hernquist}}]{Sijacki07}
{Sijacki}, D., {Springel}, V., {Di Matteo}, T., \& {Hernquist}, L. 2007,
  \mnras, 380, 877

\bibitem[{{Singh} \& {Sharma}(2015)}]{Singh15}
{Singh}, A., \& {Sharma}, P. 2015, \mnras, 446, 1895

\bibitem[{{Soker}(2003)}]{Soker03}
{Soker}, N. 2003, \mnras, 342, 463

\bibitem[{Spitzer(1962)}]{Spitzer62}
Spitzer, L. 1962, Physics of Fully Ionized Gases, 2nd edn. (New York: Wiley)

\bibitem[{{Stewart} {et~al.}(1984){Stewart}, {Fabian}, {Nulsen}, \&
  {Canizares}}]{Stewart84}
{Stewart}, G.~C., {Fabian}, A.~C., {Nulsen}, P.~E.~J., \& {Canizares}, C.~R.
  1984, \apj, 278, 536

\bibitem[{{Stone} \& {Norman}(1992)}]{Stone92}
{Stone}, J.~M., \& {Norman}, M.~L. 1992, \apjl, 390, L17

\bibitem[{{Sutherland} \& {Dopita}(1993)}]{SutherlandDopita}
{Sutherland}, R.~S., \& {Dopita}, M.~A. 1993, \apjs, 88, 253

\bibitem[{{Tremblay} {et~al.}(2015){Tremblay}, {O'Dea}, {Baum},
  {et~al.}}]{Tremblay15}
{Tremblay}, G.~R., {O'Dea}, C.~P., {Baum}, S.~A., {et~al.} 2015, \mnras, 451,
  3768

\bibitem[{Turner(1973)}]{Turner73}
Turner, J.~S. 1973, Buoyancy Effects in Fluids (Cambridge: Cambridge Univ.\
  Press)

\bibitem[{{Vazza} {et~al.}(2013){Vazza}, {Br{\"u}ggen}, \& {Gheller}}]{Vazza13}
{Vazza}, F., {Br{\"u}ggen}, M., \& {Gheller}, C. 2013, \mnras, 428, 2366

\bibitem[{{Vernaleo} \& {Reynolds}(2006)}]{Vernaleo06}
{Vernaleo}, J.~C., \& {Reynolds}, C.~S. 2006, \apj, 645, 83

\bibitem[{{Vogt} \& {En{\ss}lin}(2003)}]{Vogt03}
{Vogt}, C., \& {En{\ss}lin}, T.~A. 2003, \aap, 412, 373

\bibitem[{{Voigt} \& {Fabian}(2004)}]{Voigt04}
{Voigt}, L.~M., \& {Fabian}, A.~C. 2004, \mnras, 347, 1130

\bibitem[{{Voit}(2011)}]{Voit11}
{Voit}, G.~M. 2011, \apj, 740, 28

\bibitem[{{Voit} {et~al.}(2015){Voit}, {Donahue}, {Bryan}, \&
  {McDonald}}]{Voit15}
{Voit}, G.~M., {Donahue}, M., {Bryan}, G.~L., \& {McDonald}, M. 2015, \nat,
  519, 203

\bibitem[{{Wagh} {et~al.}(2014){Wagh}, {Sharma}, \& {McCourt}}]{Wagh14}
{Wagh}, B., {Sharma}, P., \& {McCourt}, M. 2014, \mnras, 439, 2822

\bibitem[{{Wagner} {et~al.}(2012){Wagner}, {Bicknell}, \& {Umemura}}]{Wagner12}
{Wagner}, A.~Y., {Bicknell}, G.~V., \& {Umemura}, M. 2012, \apj, 757, 136

\bibitem[{{Werner} {et~al.}(2013){Werner}, {Oonk}, {Canning}, {Allen},
  {Simionescu}, {Kos}, {van Weeren}, {Edge}, {Fabian}, {von der Linden},
  {Nulsen}, {Reynolds}, \& {Ruszkowski}}]{Werner13}
{Werner}, N., {Oonk}, J.~B.~R., {Canning}, R.~E.~A., {Allen}, S.~W.,
  {Simionescu}, A., {Kos}, J., {van Weeren}, R.~J., {Edge}, A.~C., {Fabian},
  A.~C., {von der Linden}, A., {Nulsen}, P.~E.~J., {Reynolds}, C.~S., \&
  {Ruszkowski}, M. 2013, \apj, 767, 153

\bibitem[{{Werner} {et~al.}(2014){Werner}, {Oonk}, {Sun}, {et~al.}}]{Werner14}
{Werner}, N., {Oonk}, J.~B.~R., {Sun}, M., {et~al.} 2014, \mnras, 439, 2291

\bibitem[{{Wilman} {et~al.}(2005){Wilman}, {Edge}, \& {Johnstone}}]{Wilman05}
{Wilman}, R.~J., {Edge}, A.~C., \& {Johnstone}, R.~M. 2005, \mnras, 359, 755

\bibitem[{{Yang} {et~al.}(2012{\natexlab{a}}){Yang}, {Ruszkowski}, {Ricker},
  {Zweibel}, \& {Lee}}]{Yang12}
{Yang}, H.-Y.~K., {Ruszkowski}, M., {Ricker}, P.~M., {Zweibel}, E., \& {Lee},
  D. 2012{\natexlab{a}}, \apj, 761, 185

\bibitem[{{Yang} {et~al.}(2013){Yang}, {Ruszkowski}, \& {Zweibel}}]{Yang13}
{Yang}, H.-Y.~K., {Ruszkowski}, M., \& {Zweibel}, E. 2013, \mnras, 436, 2734

\bibitem[{{Yang} {et~al.}(2012{\natexlab{b}}){Yang}, {Sutter}, \&
  {Ricker}}]{Yang12b}
{Yang}, H.-Y.~K., {Sutter}, P.~M., \& {Ricker}, P.~M. 2012{\natexlab{b}},
  \mnras, 427, 1614

\bibitem[{{Zakamska} \& {Narayan}(2003)}]{Zakamska03}
{Zakamska}, N.~L., \& {Narayan}, R. 2003, \apj, 582, 162

\end{thebibliography}


\appendix
\section{A. Dependence of HBI growth on magnetic coherence length}
\label{appendix}

\begin{figure}[tbp]
\begin{center}
\includegraphics[scale=0.55]{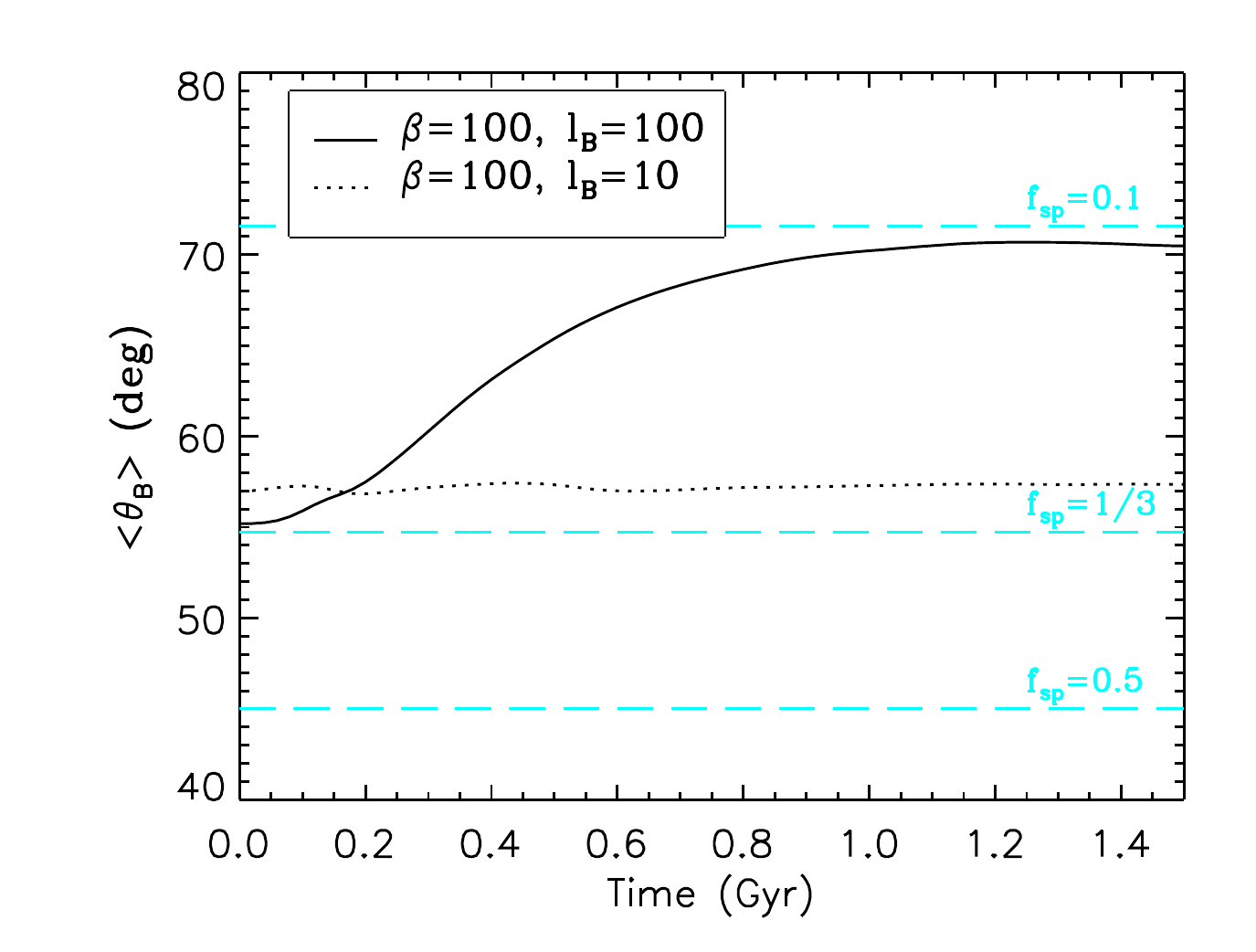} 
\caption{Same as Figure \ref{fig:thetaB} but for a cluster with a longer initial cooling time compared to Run C1 and C3. The cooling catastrophe occurs at $t \sim 2$\ Gyr. Therefore, the difference in the growth of HBI for these two cases is solely due to the different magnetic coherence length $l_{\rm B}$.}
\label{fig:thetaB_tc}
\end{center}
\end{figure}

In \S~\ref{sec:noAGN}, we showed that the HBI should not grow at all if the magnetic coherence length is smaller than the critical wavelength below which HBI is suppressed by magnetic tension, i.e., $l_{\rm B} < l_{\rm crit}$ (Eq.\ \ref{eq:lcrit}). We verified this hypothesis by comparing simulations with different $l_{\rm B}$ in the Perseus-like cluster with $\beta = 100$ (Run C1 and C3). However, as shown in Figure \ref{fig:thetaB}, it is not obvious whether the difference in these two cases is due to $l_{\rm B}$ or the fact that the magnetic field lines are stretched radially due to cooling catastrophe after $t \sim 0.3$\ Gyr (this effect is more prominent for Run C3). To this end, we ran another set of simulations with identical parameters as in Run C1 and C3, but using a cluster with a lower concentration parameter of $c=4$ and thus a longer initial cooling time of $\sim 2$\ Gyr. The result is shown in Figure \ref{fig:thetaB_tc}. Without the radial bias of the magnetic field introduced by the cooling catastrophe, it is apparent that the difference in these two runs are solely due to the different $l_{\rm B}$. That is, while the field lines become more azimuthal due to the HBI in the case where $l_{\rm B} > l_{\rm crit}$, the HBI is completely inhibited when $l_{\rm B} < l_{\rm crit}$, confirming our conclusions in \S~\ref{sec:noAGN}.


\end{document}